\documentclass[aps, pra, floatfix, twocolumn, 10pt]{revtex4-2}

\usepackage{graphicx}
\usepackage{amsfonts}
\usepackage{amsmath}
\usepackage{helvet}
\usepackage{multirow}
\usepackage{makecell}
\usepackage{bm}
\usepackage{multirow}
\usepackage{siunitx}
\usepackage[hidelinks]{hyperref}
\usepackage{booktabs}
\usepackage{chngcntr}
\usepackage{float}
\usepackage{bibunits}

\setcitestyle{super}

\DeclareSIUnit{\calorie}{cal}
\DeclareSIUnit\bar{bar}

\newcommand{\SupInfo}{}

\begin{document}

\preprint{APS/123-QED}

\newcommand{\titleinfo}{Navigating Chemical Space: Multi-Level Bayesian Optimization with Hierarchical Coarse-Graining}%
\newcommand{\uniHeidelberg}{Institute for Theoretical Physics, Heidelberg University, 69120 Heidelberg, Germany}%
\newcommand{\uniiwrHeidelberg}{Institute for Theoretical Physics and Interdisciplinary Center for Scientific Computing (IWR), Heidelberg University, 69120 Heidelberg, Germany}
\title{\titleinfo}%
\author{Luis J.~Walter}%
\affiliation{\uniHeidelberg}%
\author{Tristan Bereau}%
\email{bereau@uni-heidelberg.de}
\affiliation{\uniiwrHeidelberg}%

\date{July 15, 2025}

\begin{abstract}
Molecular discovery within the vast chemical space remains a significant challenge due to the immense number of possible molecules and limited scalability of conventional screening methods. To approach chemical space exploration more effectively, we have developed an active learning-based method that uses transferable coarse-grained models to compress chemical space into varying levels of resolution. By using multiple representations of chemical space with different coarse-graining resolutions, we balance combinatorial complexity and chemical detail. To identify target compounds, we first transform the discrete molecular spaces into smooth latent representations. We then perform Bayesian optimization within these latent spaces, using molecular dynamics simulations to calculate target free energies of the coarse-grained compounds. This multi-level approach effectively balances exploration and exploitation at lower and higher resolutions, respectively. We demonstrate the effectiveness of our method by optimizing molecules to enhance phase separation in phospholipid bilayers. Our funnel-like strategy not only suggests optimal compounds but also provides insight into relevant neighborhoods in chemical space. We show how this neighborhood information from lower resolutions can guide the optimization at higher resolutions, thereby providing an efficient way to navigate large chemical spaces for free energy-based molecular optimization.
\end{abstract}

\maketitle

\begin{bibunit}[apsrev4-2]
\section{Introduction}
\label{sec:introduction}
All molecules consist of a limited set of atoms, but their diverse properties arise from the intricate arrangements of these atoms.
The vast combinatorial possibilities of such arrangements define the so-called chemical space (CS).\citep{Kirkpatrick2004}
Exploring this space to discover new molecules with desired properties is challenging due to its immense size and complexity.\citep{Reymond2015, Polishchuk2013}
Traditionally, experimental high-throughput screening is conducted on a small subset of molecular structures to identify candidates with the desired properties.
However, this approach is costly and limited by the size of the molecular library.\citep{Mishra2008, Macarron2011}

To address these challenges, computational methods have been employed to replace expensive experiments.\citep{Fara2006}
In particular, molecular dynamics (MD) simulations can be utilized to predict the behavior of molecules based on their structure and empirical force fields.\citep{Karplus1990, Hansson2002, Bereau2021}
Combined with automated, high-throughput setups, they enable the screening of large numbers of molecules.\citep{Stanley2015}
While such simulations can reduce the cost of evaluating molecules for their target properties, they do not inherently facilitate navigation of the vast chemical search space.

Active learning methods---particularly Bayesian optimization (BO)---offer an efficient way to identify promising molecules from the extensive candidate pool.
These methods optimize functions where gradient-based approaches are inapplicable.\citep{Kushner1964, Frazier2018}
As molecular structure-property relationships generally lack gradient information, BO offers a more efficient alternative to uniform or random sampling of molecular space.\citep{Agarwal2021, Thompson2022, CrivelliDecker2024}
Since BO relies on a covariance function over the input space, a numerical representation of discrete CS is typically used to quantify molecular similarity.
For example, autoencoder models can encode molecules into latent representations.\citep{Kipf2016, GomezBombarelli2018, Reiser2022}
In contrast to fingerprint methods,\citep{Morgan1965, Warmuth2003, Muegge2015, Gorantla2024} they do not require a manual feature selection.
Although BO helps select promising candidates, it does not reduce the complexity of CS.

\begin{figure*}[t]
    \centering
    \includegraphics[width=\linewidth]{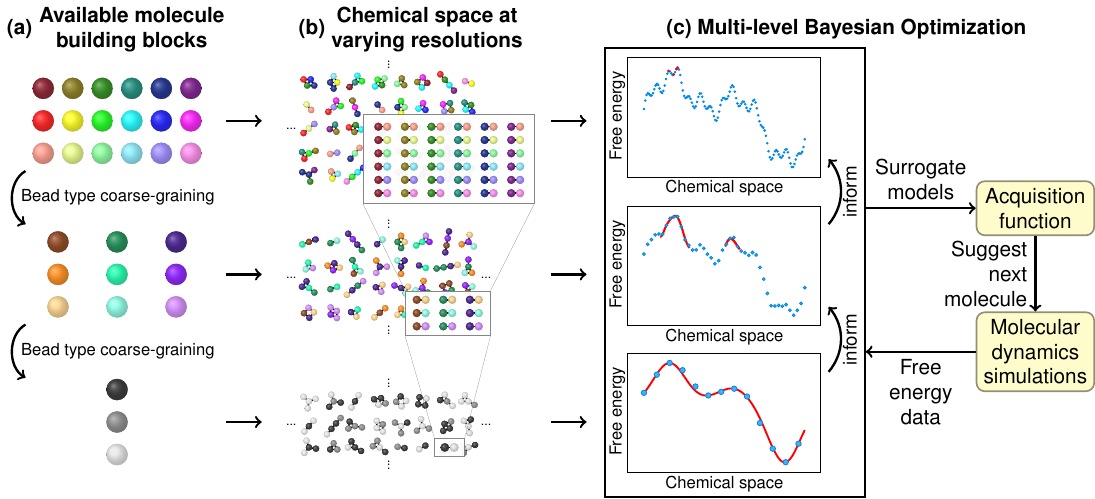}
    \caption{Overview of our multi-resolution coarse-graining molecule optimization workflow.
    (a) Definition of multiple coarse-grained (CG) models at varying resolutions. These models share the same atom-to-bead mapping but differ in bead-type assignments, with higher resolutions featuring more bead types to capture finer chemical details.
    (b) Enumeration of chemical space (CS) at different resolution levels. Higher-resolution molecules can be hierarchically mapped to lower resolutions.
    (c) Multi-level Bayesian optimization integrating information from all CS resolutions. Molecules are iteratively suggested by an acquisition function and evaluated through molecular dynamics (MD) simulations. The optimization progressively shifts toward higher-resolution evaluations. Optimization at higher-resolution levels is guided by surrogate models at lower resolutions, improving efficiency and accelerating the search for optimal candidates.}
    \label{fig:method-pipeline}
\end{figure*}

Coarse-graining---grouping atoms into pseudo-particles or beads---addresses this complexity by effectively compressing CS.
While traditionally employed to accelerate MD simulations, mapping atoms to beads reduces information and results in many-to-one relationships between atomistic and coarse-grained (CG) structures.\citep{Noid2013, Bereau2021}
The collective properties of the underlying chemical fragments determine the interactions between the CG beads.
Discretizing these interactions enables the use of a transferable CG force field, i.e., a fixed set of interaction or bead types that can be reused across the entire CS.\citep{Souza2021}
The interaction resolution of such transferable force fields, determined by the number of available CG bead types, directly impacts the many-to-one relationship between atomistic and CG structures and therefore the combinatorial complexity of CG CS.\citep{Kanekal2019}
Lower-resolution CS representations with fewer available bead types are easier to explore, but the resulting molecular structures lack detailed information.\citep{Mohr2022}
Higher resolutions provide more detailed results, but their CS representations are more challenging to explore.
This raises the question of how to combine different coarse-graining resolutions to efficiently explore CS while obtaining detailed molecular results.

In this work, we propose a multi-level BO framework for an efficient exploration of small molecule CS across multiple CG force-field resolutions.
Our method combines the reduced complexity of CS exploration at lower resolutions with a detailed optimization at higher resolutions.
The Bayesian approach provides an intuitive way to combine information from different resolutions into the optimization.
Our method builds upon the work of \citeauthor{Mohr2022}, who applied BO in a single, relatively low-resolution CG representation of CS to derive molecular design rules.\citep{Mohr2022}
They also conducted optimization in a learned representation of an enumerated CG CS. We build on their approach by integrating multiple CG resolutions into a unified optimization framework.

Our multi-level BO is related to previous multi-fidelity BO efforts,\citep{Huang2006, Fare2022, Mikkola23, Gantzler2023} which rely on different evaluation costs and accuracies for each fidelity.
In contrast, we assume a constant evaluation cost at all levels and instead utilize the varying complexity of our different CG resolutions.

Compared to recently popular generative methods for inverse molecular design,\citep{SanchezLengeling2018} our multi-level BO framework is data efficient and requires no prior training data for the optimization target.

As a demonstration of our method, we optimize a small molecule to promote phase separation in a ternary lipid bilayer.
Previous studies\citep{Barnoud2014, Centi2020} have shown that molecules embedded within lipid bilayers can modulate their phase behavior.
We quantify this phase separation behavior as a free-energy difference, which serves as the objective function for our molecular optimization.
We demonstrate that our multi-level BO algorithm effectively identifies relevant chemical neighborhoods and outperforms standard BO applied at a single resolution level.
Our proposed approach is versatile and applicable to a broad range of small-molecule optimization tasks where the target property can be expressed as a free-energy difference.

\section{Methods}
\label{sec:methods}

\subsection{Overview}
\label{sec:methods:overview}
We begin by providing an overview of our computational screening methodology.
First, we defined multiple CG models with varying resolutions, all using the same atom-to-bead mapping but differing in the assignment of transferable bead types.
Higher-resolution models featured more bead types, capturing finer chemical details while still reducing the combinatorial complexity of CS compared to the atomistic level (Figure~\ref{fig:method-pipeline}a).
This reduction allowed us to enumerate all possible CG molecules corresponding to a specific region of CS at each resolution.
Due to the hierarchical model design, higher-resolution molecules could be systematically mapped to lower resolutions (Figure~\ref{fig:method-pipeline}b).

For the next step of our molecule optimization, we embedded the CG structures into a continuous latent space using a graph neural network (GNN)-based autoencoder, with each resolution encoded separately.
This encoding step provided a smooth representation of CS, ensuring a meaningful similarity measure necessary for the subsequent BO.

Finally, a multi-level Bayesian optimization was performed based on all previously encoded CS resolutions.
The ground truth values, i.e., the optimization targets, were obtained from MD simulation-based free-energy calculations (Figure~\ref{fig:method-pipeline}c).
In our example application, such a free-energy estimate characterized the phase separation behavior of a molecule inserted into a ternary lipid bilayer.
The following sections describe each of the molecular discovery steps in detail.

\subsection{Multi-resolution Coarse-graining of CS}
\label{sec:methods:multi-resolution-cg}
Coarse-graining of molecules generally consists of two steps.
First, groups of atoms are mapped to pseudo-particles or beads.
Second, the interactions between these beads are defined based on their underlying atomistic fragments.
For both steps, the resolution of the coarse-graining can be varied.
Assigning larger groups of atoms to single beads results in a lower CG resolution for the mapping step.
Interactions between beads can be defined for each bead pair\citep{Izvekov2005, Moore2014} or discretized into a limited number of transferable bead types.
The number of available bead types then defines the interaction resolution.
Various CG models with different approaches to the mapping, discretization, and assignment of bead types exist.\citep{Pulawski2016, Ziba2019}

Since coarse-graining reduces information, a single CG molecule corresponds to multiple atomistic conformations or chemical compositions.
The CG resolution determines how many atomistic structures correspond to a single CG molecule.
Representing CS at a lower CG resolution results in fewer combinatorial possibilities for molecules and therefore a smaller CS.\citep{Kanekal2019}

We started the molecule discovery process by directly defining small molecule CS at the high-resolution CG level.
To do this, we specified the set of available CG bead types based on the relevant elements and chemical fragments from atomistic CS (Figure~\ref{fig:method-pipeline}a).
We used three CG resolution levels for our application.
They shared the same mapping of atoms to beads, but differed in the number of available bead types.
Our high-resolution model corresponded to the Martini3 model,\citep{Souza2021} a versatile CG force field with demonstrated relevance to materials design.\citep{Alessandri2021, Kjlbye2022, Mohr2022}
For our model, we ignored Martini3 bead labels, e.g., for hydrogen bonding or polarizability.
Further excluding water and divalent ions resulted in a model with 32 bead types per bead size, or 96 bead types in total.
The relationship between bead types at different resolutions was hierarchical, meaning that higher-resolution bead types could be uniquely mapped to lower resolutions.
In practice, lower-resolution bead types were obtained by averaging the interactions of higher-resolution bead types.
For the medium- and low-resolution models, we derived 45 and 15 bead types, respectively.
Section~\ref{si:derivation-lower-resolution-cg-models}{\SupInfo} provides further details on the derivation of lower-resolution models.

For all resolutions, we enumerated all possible CG molecules based on the available bead types and the defined molecule size limit of up to four CG beads (Figure~\ref{fig:method-pipeline}b).
By directly generating molecules at the CG level, the atomistic resolution was bypassed.
Since we assumed bead size-dependent but constant bond lengths and no angle or dihedral interactions, the enumeration of molecules is equivalent to the enumeration of graphs.
The small molecule size justified the neglected angle and dihedral interactions.
For the three levels of resolution, we obtained chemical spaces of approximately 90,000, \SI{6.7}{million}, and \SI{137}{million} molecules, respectively.
Section~\ref{si:molecular-graph-enumeration}{\SupInfo} elaborates details on the graph enumeration.

\subsection{Chemical Space Encoding}
\label{sec:methods:chemical-space-encoding}
From the enumeration step, we obtained large sets of molecular graphs.
While direct optimization in graph space is possible (e.g., via evolutionary algorithms\citep{vanHilten2023, Methorst2024, Luetge2025}), a numerical representation facilitates exploration of CS by enabling distance-based similarity measures.
Molecular fingerprints are often used for this purpose\citep{Morgan1965, Warmuth2003, Muegge2015, Gorantla2024} but require manual feature selection.
Instead, we used a learned projection of CS into a low-dimensional, smooth numerical representation.

For the learned encoding, we used a regularized autoencoder (RAE),\citep{Ghosh2020} which offers deterministic behavior compared to the more common variational autoencoder (VAE) architecture.\citep{Kingma2013, Kipf2016}
As we only aimed for a smooth embedding, the stochasticity of a VAE was not needed.
The built-in regularization of the RAE ensured a well-structured latent space.\citep{Ghosh2020}
We used a GNN for the node-permutation invariant encoder,\citep{Gilmer2017, Hamilton2017} which mapped molecular graphs to the five-dimensional latent space.
A decoder, composed of fully connected layers, was used to reconstruct node features and the adjacency matrix.
Although the decoder was not invariant to node permutations, the reconstruction loss ensured an invariant training of the RAE.

Input and reconstruction node features included bead-type class, size, charge, and octanol-water partition coefficient. The latter was added as a continuous feature to improve latent space structure.

We trained separate RAEs for each CG resolution using the complete set of enumerated molecules.
The separated training resulted in lower reconstruction losses and better adaptation to the reduced resolution at lower levels.
The loss combined cross-entropy terms for categorical features, a binary cross-entropy for the adjacency matrix, and a mean squared error term for the octanol-water partition coefficient.
After training, we retained only the encoder for embedding molecules.
The RAE was implemented using the PyTorch and PyTorch Geometric libraries,\citep{Paszke2019, Fey2019} following the architecture of \citeauthor{Mohr2022}\citep{Mohr2022}
Further details on the RAE architecture, the training, and an analysis of the learned latent space are provided in Section~\ref{si:rae-graph-embedding} and \ref{si:latent-space-analysis}{\SupInfo}.
In the following steps, we performed BO in these learned latent spaces.

\subsection{Single-Level Bayesian Optimization}
\label{sec:methods:single-level-bo}
Before introducing our multi-level BO approach, we first provide an overview of standard BO and our notation (see, e.g., \citeauthor{Frazier2018}\citep{Frazier2018} for a more detailed introduction).
We then describe how we extend this approach to combine multiple resolution levels into a single optimization process.
BO aims to optimize a black-box function $f: \mathcal{X} \rightarrow \mathbb{R}$ that is expensive to evaluate and has no analytical form or gradient information available. The objective is to find the global optimum $x^* = \arg \min_{x\in\mathcal{X}} f(x)$ or $x^* = \arg \max_{x\in\mathcal{X}} f(x)$ with as few function evaluations as possible.
Typically, a Gaussian process (GP) is used as a probabilistic model for $f(x)$, i.e., $f(x)\sim\mathcal{GP}(m(x), k(x,x'))$, defining a multivariate normal distribution with mean function $m(x)$ and a covariance function $k(x,x')$. This covariance kernel quantifies correlations over $\mathcal{X}$.
Although various kernel functions exist, a common choice is the radial basis function (RBF) kernel, defined as
\begin{equation}
\label{equ:rbf-kernel}
    k(x, x') = \exp\left(-\frac{1}{2\xi^2}||x - x'||^2\right),
\end{equation}
where $\xi$ is the length scale parameter.
Given training data $\mathcal{D}=\{(x_i,y_i)\}_{i=1}^n$ with inputs $X=\{x_1, \dots, x_n\}$ and observations $Y=\{y_1, \dots, y_n\}$, the posterior GP provides a predictive mean $\mu(x)$ and variance $\sigma^2(x)$ for any $x\in\mathcal{X}$.
The mean and variance are given by
\begin{equation}
\label{equ:gp-mean}
    \mu(x) = m(x) + k(x, X)K^{-1}(Y - m(X)),
\end{equation}
\begin{equation}
\label{equ:gp-std}
    \sigma^2(x) = k(x, x) - k(x, X)K^{-1}k(X, x),
\end{equation}
where $K={k(X,X)+\sigma\textsubscript{n}\mathbb{I}}$ is the covariance matrix of $X$ with an added noise term $\sigma\textsubscript{n}$.

In BO, the GP model is iteratively updated with new evaluations of the target function.
First, the function is evaluated at a set of initialization points.
Subsequent evaluations are selected based on the predictive mean and variance of the GP, guided by an acquisition function that balances exploration and exploitation.
A common choice for the acquisition function is the expected improvement (EI),\citep{Jones1998} which for minimization is defined as
\begin{equation}
    \text{EI}(x) = \mathbb{E}_{z\sim\mathcal{N}(\mu(x),\sigma^2(x))}[\max(y^* - z, 0)]
\end{equation}
with $y^* = \min_{y\in Y}y$.
The next evaluation point is determined by $x_{n+1} = \arg\max_{x \in \mathcal{X}} \text{EI}(x)$.
This process repeats until the evaluation budget is exhausted or a sufficiently good solution is found.

\subsection{Multi-Level Bayesian Optimization}
\label{sec:methods:multi-level-bo}
For our multi-level BO approach, we considered $d=3$ CG resolution levels of CS.
At each level $l\in\{1,\dots,d\}$, we defined the mapping of chemical space $\mathcal{X}_l$ to the target free-energy difference $y$ as an unknown function $f_l(x)$.
Our goal was to identify molecules at the highest resolution $d$ that are near the optimum, i.e., $x^* = \arg\min_{x \in \mathcal{X}_d} f_d(x)$, while leveraging information from the lower-resolution models ($l < d$).
Similar to the work of \citeauthor{Huang2006}, we assumed that each function $f_l(x)$ can be modeled as a correction to the lower resolution
\begin{equation}
    f_l(x) = f_{l-1}(x) + \delta_l(x),
\end{equation}
where $\delta_l(x)$ represents the correction term.\citep{Huang2006}
The hierarchical bead-type resolutions justified this delta learning approach.
We modeled each $\delta_l(x)$ as a GP, i.e.,
\begin{equation}
    \delta_l(x) \sim \mathcal{GP}(0, k_l(x, x')).
\end{equation}
with a mean function equal to zero for all $x$.
For all levels, we used an RBF kernel function (equation \ref{equ:rbf-kernel}) with level-specific length scale parameters $\xi_l$.
By definition of the GP (see equations \ref{equ:gp-mean} and \ref{equ:gp-std}), this delta learning approach corresponds to a GP with a mean prior $m(X)$ equal to the next-lower resolution function $f_{l-1}(x)$.
Thus, we can rewrite the GP for $f_l(x)$ as
\begin{equation}
\label{eq:gp-with-prior}
    f_l(x) \sim \mathcal{GP}(f_{l-1}(x), k_l(x, x')).
\end{equation}
At the lowest resolution $l=1$, no lower-level prior was available.
Instead of using a zero prior for $f_1(x)$, we applied a simple model $f_0(x)$ that approximates the free-energy difference of a molecule as the sum of the individual bead free energies.

Until now, we assumed the latent spaces of the different resolutions to be compatible.
However, since they were obtained from separate autoencoder trainings, we could not directly use a lower level function $f_l(x)$ as the prior for the GP on level $l$.
Instead, a function $\mathcal{M}_l(x)$ was required that maps points in latent space $\mathcal{X}_l$ to points in latent space $\mathcal{X}_{l-1}$.
We determined this mapping from one resolution to a lower one from the known relationships between molecules at different resolutions.
Effectively, we had a many-to-one mapping from $\mathcal{X}_l$ to $\mathcal{X}_{l-1}$, which made the mapping $\mathcal{M}_l(x)$ an unambiguous function.
Applying this mapping to equation \ref{eq:gp-with-prior}, we get
\begin{equation}
    f_l(x) \sim \mathcal{GP}(f_{l-1}(\mathcal{M}_l(x)), k_l(x, x')).
\end{equation}
as the final probabilistic model for resolution $l$.

The optimization procedure started at the lowest-resolution level $l=1$, with initialization molecules selected through weighted $k$-medoid clustering of the fully encoded CS.
The clustering weights were based on the prior of the lowest resolution and calculated as $w_i = \exp(-f_0(x_i))$.

The length scale parameters $\xi_l$ of the RBF kernels were optimized for each level using the GP marginal likelihood.
The kernel noise term $\sigma\textsubscript{n}$ (from $K$ in equations \ref{equ:gp-mean} and \ref{equ:gp-std}) was fixed to the standard deviation of the calculated free-energy differences.
This standard deviation was determined by multiple repeated evaluations of the same molecule (see Section \ref{si:free-energy-standard-deviation}{\SupInfo}).
The multi-level BO implementation used the GPyTorch library.\citep{Gardner2018}

Although BO is also possible with a batched evaluation of multiple points,\citep{Mohr2022} we only evaluated one point, i.e., one molecule, at a time.
Since each evaluation involved multiple MD simulations, we parallelized over these simulations.
We used the EI as the acquisition function on each level.
For higher levels $l>1$, the EI was computed and maximized only over CS regions with expected significant negative free-energy differences.
These regions were defined as the neighborhoods of points with promising prior information from the lower level.
Restricting the EI calculation to these neighborhoods focused the optimization on the most relevant CS regions and accelerated the EI maximization process.
Details regarding the mapping of points between latent spaces and the calculation of neighborhoods are provided in Section~\ref{si:latent-space-mapping-and-neighborhoods}{\SupInfo}.

Our multi-level BO algorithm transitions to a higher resolution when the prediction error of the GP remains below a predefined threshold for multiple consecutive evaluations.
This prediction error serves as a measure of the GP model's convergence.
For our application, we empirically set the prediction error threshold to \SI{0.12}{\kilo\calorie\per\mol} and required three consecutive evaluations below this threshold to trigger the switch.
These hyperparameters control the trade-off between exploration at lower resolutions and faster exploitation of promising regions at higher resolutions.
Lowering the threshold and increasing the number of required evaluations enhances exploration at lower resolutions, but increases the total number of molecule evaluations needed.

In addition to increasing the resolution level, the algorithm can switch back to the previous lower resolution.
Since we want to effectively leverage lower-resolution models, we are only interested in high-resolution evaluations in regions where a reliable prior is available.
If the candidate with the maximal EI is too far away from regions with a reliable prior from lower levels, we switch back to the previous resolution level.
Specifically, the criterion for switching to resolution level $l-1$ is defined as $||x^* - x'|| > 2\xi_l$, $\forall x'\in\{x\in\mathcal{X}_{l} | x\in X_l \vee \mathcal{M}_{l}(x)\in X_{l-1}\}$, where $X_l$ denotes the set of already evaluated points at level $l$.

\subsection{Estimating the Membrane Demixing Behavior}
\label{sec:methods:estimating-demixing}

\begin{figure}[t]
    \centering
    \includegraphics[width=\linewidth]{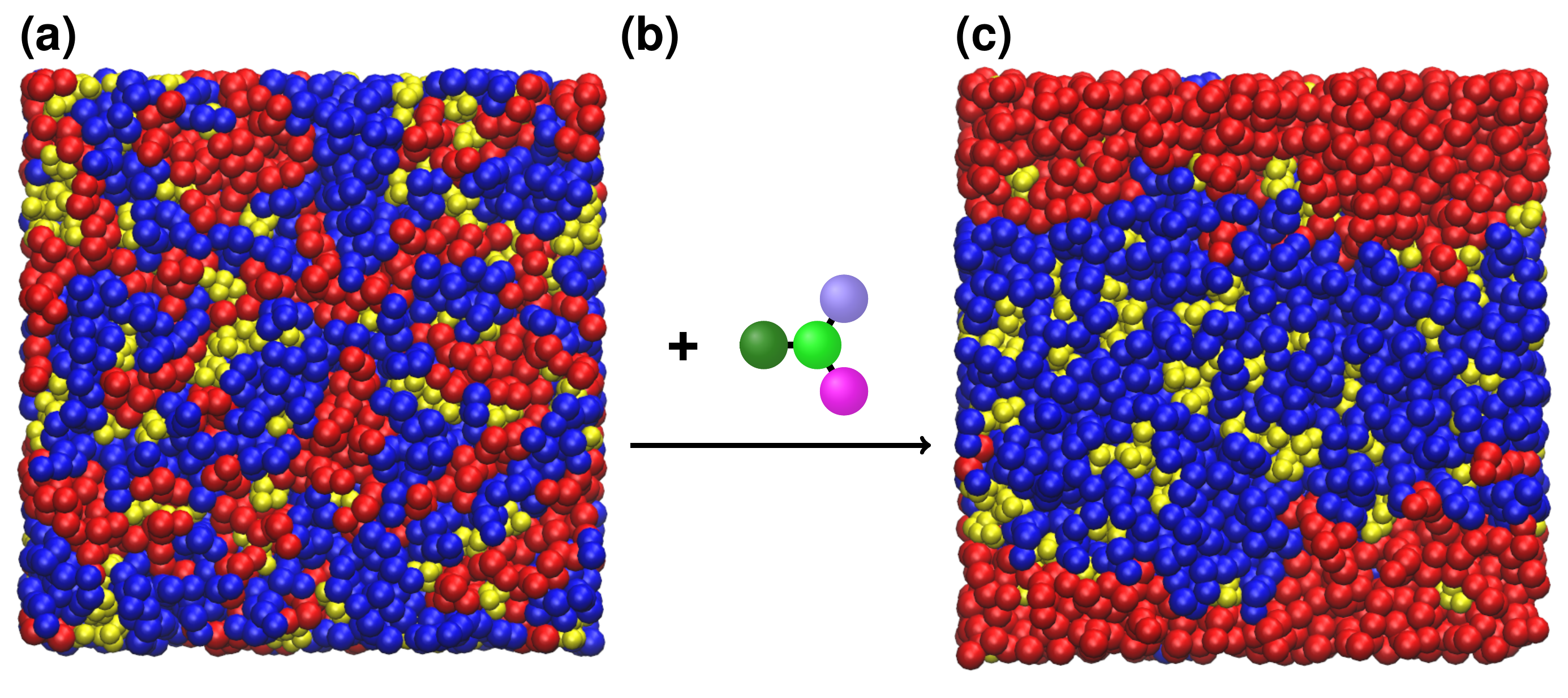}
    \caption{Influencing phase separation in a lipid bilayer by inserting small molecules. Shown is a top view of a CG ternary lipid bilayer composed of DPPC (blue), DLiPC (red), and cholesterol (yellow). (a) In the mixed state, the bilayer contains small, dispersed lipid patches. (b) Upon inserting specific small molecules, (c) the bilayer transitions to a demixed state with pronounced phase separation between the two phospholipids.}
    \label{fig:membrane-demixing}
\end{figure}

For our application, we optimized small molecules to enhance phase separation in a ternary lipid bilayer consisting of 1,2-dipalmitoyl-\textit{sn}-glycero-3-phosphocholine (DPPC), 1,2-dilinoleoyl-\textit{sn}-glycero-3-phosphocholine (DLiPC), and cholesterol (Figure~\ref{fig:membrane-demixing}).
DPPC and DLiPC differ only in their acyl chains, with DPPC having two saturated 16-carbon chains and DLiPC having two doubly unsaturated 18-carbon chains.
The phase separation can be quantified by the DPPC-DLiPC contact fraction.\citep{Barnoud2014}
However, directly observing the effect of a molecule on lipid mixing requires long simulations with large bilayer leaflets, which is impractical for high-throughput screening.
Alternatively, potential of mean force (PMF) profiles along the axis perpendicular to the bilayer plane can be compared for pure DPPC, DLiPC, and ternary bilayers.\citep{Centi2020}
Since PMF calculations (e.g., via umbrella sampling\citep{Torrie1977}) are still computationally expensive, we employed thermodynamic integration (TI) \citep{Chipot2007, Mey2020} calculations at a few key positions in the bilayers as a proxy.
\citeauthor{Centi2020} showed that molecules that influence the demixing or mixing of a DPPC-DLiPC bilayer localize near the bilayer center because the two phospholipids differ only in their carbon tails.\citep{Centi2020}
\begin{figure}[t]
    \centering
    \includegraphics[width=\linewidth]{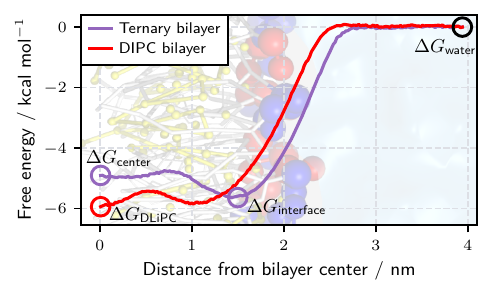}
    \caption{Estimating the demixing behavior of molecules via free-energy calculations at four bilayer depths (circles) as an alternative to potential of mean force (PMF) computations (solid lines). For the molecule optimization, we aim to minimize $\Delta\Delta G = \Delta G\textsubscript{center} - \Delta G\textsubscript{DLiPC}$ under the conditions that $\Delta G\textsubscript{center} < \Delta G\textsubscript{interface}$ and $\Delta G\textsubscript{center} < \Delta G\textsubscript{water}$. The background illustrates the hydrophobic tails (grey), the charged headgroups of DPPC (blue) and DLiPC (red), as well as cholesterol (yellow).
    The plotted PMFs correspond to a molecule with $\Delta G\textsubscript{center} > \Delta G\textsubscript{interface}$, indicating that it localizes at the bilayer interface and therefore does not significantly influence lipid mixing.}
    \label{fig:pmf-points}
\end{figure}
To determine a molecule's preferred localization, we performed TI computations at the center ($z=\SI{0}{\nano\meter}$) of the ternary bilayer, at the interface ($z=\SI{1.5}{\nano\meter}$) and in bulk water (Figure~\ref{fig:pmf-points}), obtaining the free-energies $\Delta G\textsubscript{center}$, $\Delta G\textsubscript{interface}$, and $\Delta G\textsubscript{water}$, respectively.
We initially used $\Delta G\textsubscript{center}$ and $\Delta G\textsubscript{water}$ to identify non-inserting molecules, allowing us to skip further free-energy evaluations for these cases.
\citeauthor{Centi2020} showed that molecules that enhance the phospholipid demixing localize near the DLiPC phase.\citep{Centi2020}
Therefore, we performed a fourth TI calculation at the center of a pure DLiPC bilayer, yielding $\Delta G\textsubscript{DLiPC}$.
Unlike the direct observation of DPPC-DLiPC contacts, $\Delta G$-based scoring was easily parallelized, thereby further reducing the wall time per evaluated molecule.
The main optimization target was the free-energy difference, $\Delta\Delta G = \Delta G\textsubscript{center} - \Delta G\textsubscript{DLiPC}$, assuming the molecule localizes near the ternary bilayer center.
To ensure robust optimization even when molecules localize at the interface or in the water, we combined $\Delta\Delta G$ with a score $S$ defined as a conditional weighted sum of $\Delta G\textsubscript{water} - \Delta G\textsubscript{interface}$ and $\Delta G\textsubscript{interface} - \Delta G\textsubscript{center}$.
In contrast to using a constant $\Delta\Delta G$ for interface- or water-localizing molecules, this score provided a more nuanced direction for optimization, steering it toward relevant regions of CS.
Negative $\Delta\Delta G$ values indicated a preference for the DLiPC phase, corresponding to a demixing behavior.
Overall, the molecule optimization corresponded to a minimization of $\min(\Delta\Delta G, 0) + S$.
Section~\ref{si:score-calculation}{\SupInfo} provides further information on the calculation of the score $S$.

\subsection{Molecular Dynamics Simulations}
\label{sec:methods:md-simulations}
We used MD simulations in a high-throughput manner\citep{Menichetti2019, Hoffmann2020} to perform the TI calculations of the free-energy differences.
All MD simulations were performed using GROMACS 2024.2.\citep{Abraham2015, Pall2020}
Martini3 and Martini3-derived (see Section~\ref{sec:methods:multi-resolution-cg}) force fields were used for the CG simulations.\citep{Souza2021, BorgesArajo2023}
The derived lower-resolution bead types are compatible with the standard Martini3 bead types and can therefore be evaluated within unmodified Martini3 environments.

Our lipid bilayer simulation setup was based on the protocol by \citeauthor{Ozturk2024}\citep{Ozturk2024}
We used a leap-frog stochastic dynamics integrator with an integration time step of \SI{20}{\femto\second} (in reduced CG units).
All simulations were performed in the $NPT$ ensemble at a temperature of \SI{305}{\kelvin} and pressure of \SI{1}{bar},\citep{Centi2020} controlled by a semi-isotropic C-rescale barostat.\citep{Bernetti2020}
For the TI, we used 26 linearly-spaced $\lambda$ steps for the decoupling of Lennard-Jones interactions and 10 additional linear $\lambda$ steps for the decoupling of Coulomb interactions in the case of charged molecules.
Since each molecule evaluation required up to four TI calculations, each with up to 36 $\lambda$ steps, evaluating a single molecule could require up to 144 individual simulations.
Further simulation parameters are provided in Section~\ref{si:simulation-parameters}{\SupInfo}.
The package MBAR\citep{Shirts2008, Wu2024} was used to calculate free-energy differences from the MD simulation data.

Membrane systems were generated using the program insane.\citep{Wassenaar2015}
Following the approach of \citeauthor{Centi2020}, we used a lipid composition of DPPC:DLiPC:cholesterol in a 7:4.7:5 ratio.\citep{Centi2020}
For a bilayer area of $6\times\SI{6}{\nano\meter\squared}$, used for the free-energy evaluations, this corresponded to 26 DPPC, 18 DLiPC, and 19 cholesterol molecules per bilayer leaflet.
We used the colvars module\citep{Fiorin2013} in GROMACS to calculate or restrain the phospholipid contact fraction.
Specifically, the collective variable was defined as the coordination number between the first {\sffamily C1} beads of DLiPC and DPPC with a cutoff distance of \SI{1.1}{\nano\meter}.\citep{Centi2020}
During the TI simulations, the coordination number was restrained to 65 contacts per leaflet, yielding an average of 2.5 DLiPC molecules within the cutoff per DPPC.
This slightly exceeds the 2.15 contacts expected from random lipid placement by insane.\citep{Wassenaar2015}

\section{Results and Discussion}
\label{sec:results}

\subsection{Multi-level Bayesian Optimization}
\label{sec:results:multi-level-bayesian-optimization}

\begin{figure*}[ht]
    \centering
    \includegraphics[width=\linewidth]{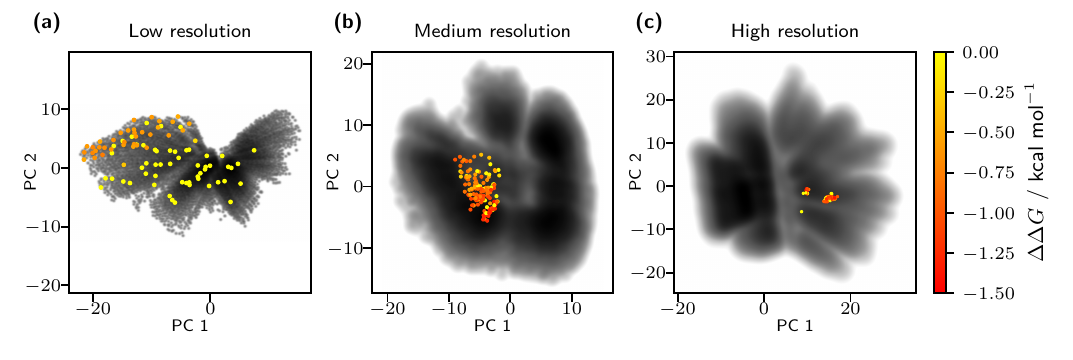}
    \caption{Encoded chemical spaces and evaluated points for the three levels of resolution. The full chemical spaces are shown as kernel-density estimations of latent spaces principal component analysis (PCA) projections (black). Evaluated molecules across the three resolutions are overlaid as colored points (yellow to red),  where lower $\Delta\Delta G$ values indicate stronger lipid bilayer demixing. Due to separate encodings at each resolution, latent space points are not directly transferable. (a) Optimization proceeds from broad, low-resolution exploration to (b) progressively focused searches in medium and (c) high resolutions.}
    \label{fig:latent-space-points}
\end{figure*}

We applied our multi-level BO workflow to identify small molecules that enhance the phase separation of a ternary lipid bilayer, demonstrating its effectiveness in navigating chemical space.
We restricted the search to small molecules with up to 16 heavy atoms, corresponding to a maximum of four beads in our CG model.
We imposed no additional constraints, such as the presence of specific functional groups, to rigorously test our method.
Our multi-level molecule optimization utilized three coarse-graining resolutions, incorporating 15, 45, and 96 distinct bead types.
While all three levels use the same spatial coarse-graining, complexity increased with the combinatorial diversity of bead types, spanning approximately 90,000, 6.7 million, and 137 million possible CG molecules.
To identify phase separation-enhancing molecules at the highest resolution, we used lower-resolution models only to guide the search, thereby reducing the complexity of the optimization compared to direct high-resolution exploration.
At all levels, a molecule's effect on phase separation was quantified by an MD simulation-derived free-energy difference, $\Delta\Delta G$ (see Section~\ref{sec:methods:estimating-demixing}).

The optimization was conducted within RAE-learned latent embedding spaces, generated from the CG models at each resolution.
As a first step, we computed the $\Delta\Delta G$ values for all 15 low-resolution bead types.
These results enabled us to construct a cost-effective prior for the low-resolution model, based on an additivity assumption over individual bead values (see Section \ref{si:additivity-assumption}{\SupInfo} for a detailed evaluation of this assumption).
Using this prior, we initialized the multi-level active learning with 50 low-resolution molecules.
Subsequent molecules and their resolution levels were determined iteratively by our multi-level BO algorithm.
We evaluated 327 molecules in total: 106 molecules (15+50+41) at the low resolution, 148 at the medium resolution, and 73 at the high resolution.
In each iteration, a single molecule was selected for evaluation using MD simulations.
The resulting $\Delta\Delta G$ value was then used to update the BO model, which informed the selection of the following molecule.

\begin{figure}[t]
\includegraphics[width=\linewidth]{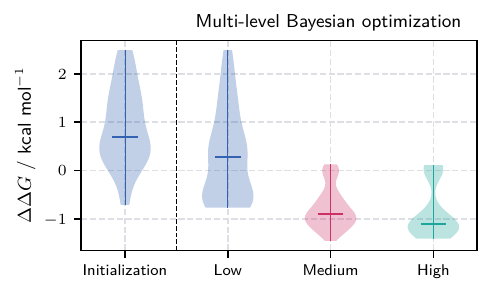}
\caption{Distribution of $\Delta\Delta G$ values for all evaluated candidates at different stages of the multi-level Bayesian optimization process. Violin plots show the distributions for the initialization set and candidates evaluated at low, medium, and high-resolution levels. As the optimization progresses to higher resolutions, the distribution of $\Delta\Delta G$ values progressively shifts toward lower (more favorable) values. Horizontal bars indicate the median of each distribution.}
\label{fig:optimization-progress-violin}
\end{figure}

Our multi-level BO approach progressively narrows the search space through the three resolution levels.
The optimization begins with a broad exploration of low-resolution CS, identifying coarse regions likely to contain molecules with favorable $\Delta\Delta G$ values.
Insights from this stage inform the medium-resolution search, allowing the algorithm to focus on more promising sub-regions.
This process is further refined at the high-resolution level to pinpoint localized areas within CS that are most likely to yield effective candidates.
By leveraging information from the preceding levels, the algorithm bypasses large areas of the CS landscape that are unlikely to yield relevant molecules.
Therefore, the number of required evaluations and the overall computational cost are reduced.
Figure~\ref{fig:latent-space-points} presents 2D projections of the encoded CS (black) together with the evaluated molecules.
Because each resolution is encoded independently, the representations differ and prevent a direct transfer of points.
However, molecules can be readily mapped across latent spaces by leveraging the known mapping between bead types.
The figure illustrates the funnel-like optimization: as resolution increases, the search becomes more focused, eventually concentrating on localized sub-regions of chemical space.
Many low-resolution candidates display unfavorable $\Delta\Delta G$ values or negligible effects on phase separation (yellow).
In contrast, searches at medium and high resolutions increasingly yield molecules with lower $\Delta\Delta G$ values corresponding to a more substantial impact on lipid demixing (orange to red).
Figure~\ref{fig:optimization-progress-violin} further illustrates this trend, showing the distribution of evaluated $\Delta\Delta G$ values across the three resolution levels, including the initialization points at resolution $l=1$.
Candidates from the low-resolution optimization already show lower $\Delta\Delta G$ values relative to the initialization set.
However, higher-resolution candidates generally exhibited even stronger phase-separation effects, with medium resolution peaking around \SI{-1}{\kilo\calorie\per\mole} and high resolution around \SI{-1.2}{\kilo\calorie\per\mole}.
The differences between the low- and medium-resolution minima support our hypothesis about the varying smoothness of the free-energy landscape across resolutions.

The computational cost per simulation is the same across all three resolutions. Consequently, the overall computational load at each level is primarily determined by the number of evaluated molecules.
For non-inserting molecules, two of the four TI calculations can be omitted (see Section~\ref{sec:methods:estimating-demixing}). As the lowest resolution filtered out most non-inserting molecules, its average computational load per evaluation was slightly lower than at higher resolutions.

We terminated the optimization after 73 high-resolution evaluations, as no further improvement in $\Delta\Delta G$ was observed.
The 327 evaluated molecules correspond to less than \SI{3e-4}{\percent} of the total high-resolution molecule space.
While global optimality is not guaranteed, the workflow identified multiple promising candidates with pronounced effects on lipid phase separation despite limited evaluations.

\subsection{Evaluation of Optimized Molecules}
\label{sec:results:eval-optimized-molecules}
Following the overall optimization process analysis, we now focus on the top candidate molecules with the lowest $\Delta\Delta G$ values.
As the Martini3 CG model (without bead labels)\citep{Souza2021} corresponds to our high-resolution model, the optimized molecules do not provide atomistic details but reveal valuable insights into the chemical moieties driving the phospholipid phase separation.
\begin{figure}[t]
    \includegraphics[width=\linewidth]{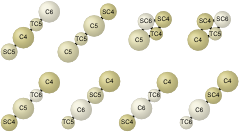}
    \caption{CG structures of the best eight high-resolution molecule candidates identified in the optimization process. The molecules exhibit low free-energy values ($\Delta\Delta G$) below \SI{-1.3}{\kilo\calorie\per\mole}, indicating a strong influence on phospholipid bilayer phase separation. All molecules are exclusively composed of hydrophobic {\sffamily C4}, {\sffamily C5}, and {\sffamily C6} beads (in different sizes, indicated by prefixes S/T), corresponding to Martini3 types for alkenes, aromatic rings, and sulfide groups, respectively. Six of the eight molecules have an extended/chain-like topology.}
    \label{fig:top-molecules}
\end{figure}
The top eight CG molecules, shown in Figure~\ref{fig:top-molecules}, all display $\Delta\Delta G$ values below \SI{-1.3}{\kilo\calorie\per\mole}, with the best candidate at \SI{-1.4}{\kilo\calorie\per\mole} (top left of the figure).
These results indicate a strong effect on the phase separation.
A consistent feature across all eight CG molecules is the exclusive presence of hydrophobic {\sffamily C4}, {\sffamily C5}, and {\sffamily C6} beads in varying bead sizes.
These Martini3 beads correspond to alkenes, aromatic rings, and thiol/sulfide groups, respectively.\citep{Souza2021}
This observation aligns with \citeauthor{Barnoud2014}, who showed that aromatic groups promote demixing, while aliphatic groups ({\sffamily C1}, {\sffamily C2}, and {\sffamily C3} beads) favor phospholipids mixing.\citep{Barnoud2014}
The two distinct topologies shown in Figure~\ref{fig:top-molecules} correspond to the two prominent point clusters in the 2D projection of Figure~\ref{fig:latent-space-points}c. While each cluster contains molecules with a variety of topologies, the highest-scoring molecules within them are predominantly of the two topologies in Figure~\ref{fig:top-molecules}.

The highest-performing molecules at both low and medium resolution (see Section~\ref{si:best-low-medium-molecules}{\SupInfo}) exhibit more diverse topologies but share similar trends in bead-type composition. While the low-resolution results already provide preliminary chemical insights, more detailed information---such as the unfavorable contribution of {\sffamily C1}, {\sffamily C2}, and {\sffamily C3} beads---only becomes evident through the inclusion of higher-resolution models.

Directly measuring bilayer phase separation requires significant simulation time and is therefore computationally expensive.
Instead, we estimated demixing effects from free-energy differences.
To validate this approach and confirm that the identified candidates indeed promote phase separation, we perform \SI{1600}{\nano\second} MD simulations (in reduced CG units) of the best candidate (top left in Figure~\ref{fig:top-molecules}) in a ternary lipid bilayer system.
Using this method to evaluate the demixing effect required one to two orders of magnitude more wall time than the free energy-based scoring used for the optimization.
As a reference, we employ benzene, previously identified by \citeauthor{Barnoud2014} as a potent driver of lipid bilayer phase separation.\citep{Barnoud2014}
Following their protocol, we use a solute/lipid mass ratio of 4.8\% (see Section~\ref{si:demixing-simulation-composition}{\SupInfo} for composition details).
Phase separation was quantified by tracking DLiPC and DPPC contacts over the simulation trajectory.
\begin{figure}[t]
    \includegraphics[width=\linewidth]{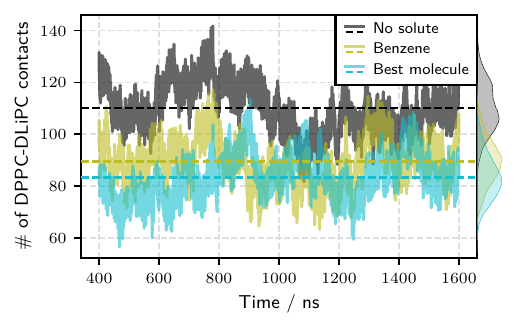}
    \caption{Time evolution of DPPC-DLiPC lipid contacts in ternary bilayers over \SI{1200}{\nano\second} CG MD simulations (excluding \SI{400}{\nano\second} for equilibration). Three conditions are compared: a bilayer without solutes (black), a bilayer containing benzene as a known demixing agent (olive green), and one with the top-performing optimized molecule from Figure~\ref{fig:top-molecules} (cyan), each at a solute/lipid mass ratio of 4.8\%. Dashed horizontal lines indicate mean contact numbers. The optimized molecule reduces DPPC-DLiPC contacts more than benzene, demonstrating a stronger phase-separation effect.}
    \label{fig:contact-count-high}
\end{figure}
Figure~\ref{fig:contact-count-high} presents the evolution of these contacts throughout the simulation, with dashed lines indicating average values.
Each trajectory's initial \SI{400}{\nano\second} were discarded to ensure equilibration.
Additionally, a control simulation without any added solute was conducted.
Compared to this bilayer without solutes, our best candidate substantially reduced DLiPC-DPPC contacts, indicating a pronounced effect on bilayer demixing.
Our best candidate also outperforms benzene, producing a greater reduction in the number of contacts, suggesting a stronger influence on phospholipid phase separation.

\begin{figure}[t]
    \includegraphics[width=\linewidth]{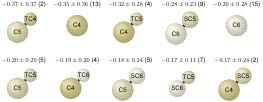}
    \caption{Top ten most influential molecular features contributing to lipid bilayer phase separation, identified via LASSO regression of the optimized CG molecules with $\Delta\Delta G < 0$. Features were limited to single beads and bead pairs. Each panel displays a feature's structure along with its corresponding regression coefficient, bootstrapped uncertainty, and frequency of occurrence within the dataset (number in parentheses). Features only involve hydrophobic {\sffamily C4}, {\sffamily C5}, and {\sffamily C6} beads and pairs of differing bead sizes.}
    \label{fig:molecule-features}
\end{figure}

To identify relevant molecular features and design rules from the set of optimized molecules, we applied LASSO regression analogous to \citeauthor{Mohr2022}\citep{Mohr2022}
Derived rules could subsequently inform the design of atomistic structures.
We analyzed single-bead and bead-pair features across all molecules with $\Delta\Delta G < 0$, yielding 85 features.
Higher-order features were not included due to the size of the dataset.
Feature extraction and LASSO regression details are provided in \citeauthor{Mohr2022}\citep{Mohr2022}
The top ten most relevant molecular features, along with their regression coefficients, bootstrapped uncertainties, and frequencies of occurrence, are shown in Figure~\ref{fig:molecule-features}.
Consistent with our earlier analysis of the top eight molecules, the most influential features involve hydrophobic {\sffamily C4}, {\sffamily C5}, and {\sffamily C6} beads.
Moreover, combinations of a regular-sized and tiny or small-sized bead (indicated by T or S) appear relevant.
These derived features provide interpretable insights into the physical interaction mechanisms that drive bilayer phase separation.
They can be used to design atomistic molecular structures that exhibit the same phase separation behavior.

\subsection{Comparison with Standard BO}
\label{sec:results:comparison-with-standard-bo}

\begin{figure}[t]
    \includegraphics[width=\linewidth]{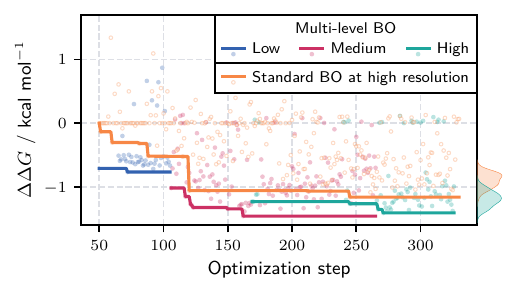}
    \caption{Progression of the $\Delta\Delta G$ values during multi-level and standard BO runs. Multi-level BO uses evaluations at low (blue), medium (magenta), and high (green) resolutions, while the standard BO (orange) operates only at the high resolution. Solid lines show the current best value during the optimization. Initialization points are excluded. The multi-level case accounts for the 15 extra evaluations for the low-resolution prior. Kernel density estimates (right edge) reflect the distribution of best 50 high resolution candidates. Multi-level BO consistently achieves lower $\Delta\Delta G$ values, as indicated by the shifted distribution.}
    \label{fig:comparison-multi-vs-single-level}
\end{figure}

Is multi-level BO computationally advantageous compared to BO using only the high-resolution model?
To address this, we performed standard BO with the same number of initial points and total evaluations as in the multi-level case.
While BO is typically benchmarked by averaging the cumulative best result across multiple runs to reduce initialization bias, this is computationally infeasible for our bilayer demixing system.
Instead, we compare the distributions of obtained $\Delta\Delta G$ values and the cumulative best result within single runs.
We provide a toy model comparison of results averaged over multiple runs in Section~\ref{si:toy-model}{\SupInfo}.
Figure~\ref{fig:comparison-multi-vs-single-level} presents the progression of the best $\Delta\Delta G$ values for both optimization approaches.
The diagram excludes the 50 initialization points and accounts for the 15 additional evaluations required to construct the low-resolution prior for the multi-level approach.
The multi-level BO consistently outperforms the standard BO, achieving superior cumulative best values (solid lines) across all resolution levels.
Additionally, the distribution (based on the best 50 molecules) and scatter plots in orange and green highlight that multi-level BO not only finds a better overall candidate, but multiple candidates with significantly lower $\Delta\Delta G$ values than the standard BO optimization.
The peak of the multi-level BO distribution is shifted toward lower $\Delta\Delta G$ values compared to the standard BO optimization.

\subsection{Chemical Neighborhood Sizes Across Resolutions}
\label{sec:results:neighborhoods}

Our multi-level BO algorithm relies on the assumption that the free-energy landscape over the learned chemical representations is smoother at lower resolutions.
To test this, we introduce the concept of \textit{chemical neighborhoods} and analyze their sizes across different resolution levels.
We define a chemical neighborhood as a region in chemical space containing similar molecules.
Similarity implies that known properties about one molecule help predict properties of its neighbors.
Here, neighborhood size is determined by the lengthscale $\xi_l$ of an RBF kernel fitted in a GP regression.
This length scale quantifies correlations between points in the latent space and is thus an intrinsic measure of chemical neighborhood size.
To obtain the $\xi_l$, we fit independent GP models to the evaluated molecules at each resolution level, excluding lower-resolution priors to prevent bias.
Neighborhood size is then calculated as the average number of neighbors within a distance $d$, where $d = \alpha\,\xi_l$ and $\alpha=0.5$ determines the required similarity for a chemical neighborhood.
\begin{figure}[t]
\includegraphics[width=\linewidth]{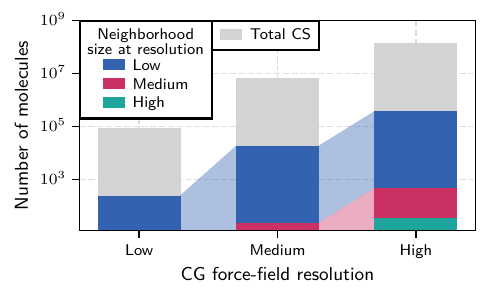}
\caption{Chemical neighborhood sizes across different CG resolutions of CS. The chart shows the number of molecules within a chemical neighborhood at low (blue), medium (pink), and high (teal) resolution, derived by fitting the lengthscale of a GP RBF kernel to the evaluated molecule data. The total number of molecules is shown in gray. Lower-resolution neighborhoods are mapped to higher resolutions by considering average molecule densities. These different neighborhood sizes reflect the varying smoothness of the free-energy landscape across different CS resolutions.}
\label{fig:neighborhood-sizes}
\end{figure}
Figure~\ref{fig:neighborhood-sizes} shows relationships between the obtained neighborhood sizes, the total number of molecules in the chemical space, and neighborhoods from lower resolutions mapped to higher resolutions (exact numbers in Section~\ref{si:chemical-neighborhood-sizes}{\SupInfo}).
Considering the logarithmic scale of the y-axis, we observe that neighborhood sizes span several orders of magnitude across the three resolutions.
When mapped to medium or high resolution, low-resolution neighborhoods with about 249 molecules expand to about 18,600 and 378,000 molecules.
Similarly, a medium-resolution neighborhood with about 23 molecules maps to a neighborhood of 468 molecules at high resolution.
This exponential scaling suggests that prior information for many high-resolution molecules can be inferred from relatively few low-resolution evaluations.
Section~\ref{si:latent-space-points-mapping}{\SupInfo} further illustrates this by showing the coverage of the higher-resolution latent spaces by mapping evaluated molecules from the lower resolutions.
These results support our assumption of a smoother free-energy landscape at lower resolutions.

\section{Conclusions}
\label{sec:conclusion}
This work introduces a multi-level Bayesian optimization (BO) framework for efficient exploration of chemical space (CS).
Our method employs multiple levels of coarse-graining to exploit the varying smoothness of free-energy landscapes across different model resolutions.
By informing the optimization process at higher resolutions with prior knowledge from lower resolutions, we accelerate the search for optimal molecules.
Our BO-based algorithm combines information from multiple resolutions in a Bayesian manner, enabling a funnel-like optimization process through CS.
This approach allows us to bypass irrelevant regions of CS at higher-resolution representations, substantially reducing the number of required molecule evaluations and the overall computational cost.
We demonstrate the effectiveness of our method by identifying small coarse-grained (CG) molecules that enhance phase separation in a ternary lipid bilayer.
Despite evaluating only approximately \SI{3e-4}{\percent} of the total number of high-resolution molecules and assuming no prior knowledge of relevant CS regions, we successfully identified several candidates with a significant impact on lipid bilayer phase separation.
Our multi-level approach outperforms standard BO, achieving a better overall best result and obtaining a significantly shifted distribution of evaluated molecules toward stronger effects on phase separation.
The optimized CG molecules enable us to extract relevant molecular features and design rules.
Our analysis of chemical neighborhood sizes at different resolutions confirms the assumption of smoother free-energy landscapes at lower resolutions.
Notably, obtained neighborhood sizes vary by several orders of magnitude, allowing us to get prior information for many molecules at high resolution from a small number of evaluations at low resolution.

In this study, we limited our funnel optimization to the CG level and thus did not derive atomistic structures for the identified candidates.
Similar to \citeauthor{Mohr2022}, atomistic structures could be reconstructed based on the extracted molecular features.\citep{Mohr2022}
Notably, these features provide an intuitive and interpretable summary of the key chemical factors, providing valuable insight into the underlying physical interaction mechanisms.
Moreover, the atomistic resolution could be integrated directly into our multi-level optimization framework.
Since each CG bead maps to $10^2$--$10^4$ atomistic fragments,\citep{Menichetti2019} the atomistic chemical space is vastly larger. Combined with evaluation costs two to three orders of magnitude higher,\citep{May2013, Alessandri2023} this poses challenges.
Nevertheless, these cost differences enable approaches like multi-fidelity BO,\citep{Huang2006, Gantzler2023} and high-resolution CG results generally provide an efficient starting point that reduces the number of required atomistic evaluations.

A limitation of our multi-level BO method is its reliance on a hierarchical relationship between resolutions, with higher resolutions required to exhibit sufficient complexity.
Although multi-level BO improves efficiency over standard BO for complex optimization landscapes, it may underperform on simpler problems.
In our application, the target function---mapping the learned latent representation of CS to free energy---is sufficiently complex and non-smooth to benefit from the multi-level BO strategy. Further work is needed to identify optimal complexity hierarchies and resolution-level differences, which could further enhance efficiency.
Another limitation is the increased complexity in implementation and hyperparameter tuning. Multi-level BO requires setting hyperparameters for each resolution, as well as additional parameters for resolution switching.
Nevertheless, these hyperparameters are primarily related to the chemical space and can thus be transferred across different molecular optimization tasks.

Beyond its demonstrated application in lipid bilayer phase separation, our multi-level BO framework can solve other optimization problems characterized by free-energy differences.
We expect our method to be particularly advantageous in applications with little prior knowledge or training data.
Furthermore, integrating our method with a FAIR\citep{Wilkinson2016} data infrastructure and automated simulation workflows, such as Martignac,\citep{Bereau2024} will enhance data management, reproducibility, and end-to-end automation, thereby making the multi-level BO approach more systematic and streamlined.

Our work provides a versatile and efficient molecular design and optimization framework, offering a promising direction for tackling complex chemical search problems.

\section*{Acknowledgments}
The authors would like to thank Daniel Nagel and Luis Itzá Vázquez-Salazar for constructive criticism of the manuscript.
T.B. acknowledges support by the Deutsche Forschungsgemeinschaft (DFG, German Research Foundation) under Germany's Excellence Strategy EXC 2181/1–390900948 (the Heidelberg STRUCTURES Excellence Cluster).
L.W. and T.B. acknowledge the SIMPLAIX project funded via the Klaus Tschira Stiftung gGmbH for its support.\\

\section*{Data availability}
The code for the multi-level Bayesian optimization workflow, the simulation setup, the analysis, and the autoencoder training, as well as the autoencoder models and free-energy results, can be found at \href{https://github.com/BereauLab/Multi-Level-BO-w-Hierarchical-CG}{https://github.com/BereauLab/Multi-Level-BO-w-Hierarchical-CG}.
A representative subset of the simulation data is available on NOMAD at \href{https://doi.org/10.17172/NOMAD/2025.05.27-1}{DOI:10.17172/NOMAD/2025.05.27-1}.
We also provide a tutorial showcasing the main concepts of this paper through a simple two-bead molecule optimization: \href{https://github.com/BereauLab/Molecule-Optimization-w-Hierarchical-CG-Tutorial}{https://github.com/BereauLab/Molecule-Optimization-w-Hierarchical-CG-Tutorial}.

\putbib[biblio]
\end{bibunit}

\onecolumngrid
\clearpage
\clearpage
\onecolumngrid
\makeatletter
\def\p@subsection{SI}
\def\p@subsubsection{SI}
\renewcommand{\thesection}{\arabic{section}}
\renewcommand{\thesubsection}{\arabic{section}.\arabic{subsection}}
\renewcommand{\@seccntformat}[1]{\csname the#1\endcsname\quad}
\renewcommand{\section}{\@startsection{section}{1}{0pt}{3.5ex}{1ex}{\raggedright\large\bfseries}}
\renewcommand{\subsection}{\@startsection{subsection}{2}{0pt}{3.5ex}{1ex}{\raggedright\large\bfseries}}
\counterwithin{table}{section}
\counterwithin{figure}{section}
\renewcommand{\thefigure}{SI\arabic{section}.\arabic{figure}}
\renewcommand{\thetable}{SI\arabic{section}.\arabic{table}}
\renewcommand{\theequation}{SI\arabic{equation}}
\setcounter{equation}{0}
\makeatother
\setcounter{section}{0}
\setcounter{page}{1}
\begin{center}
    \textbf{\Large Supplementary Information for}\\
    \textbf{\Large\textit{\titleinfo}}\\
    \vspace{0.5cm}
    Luis J.~Walter\\
    \textit{\uniHeidelberg}\\
    \vspace{0.5cm}
    Tristan Bereau\\
    \textit{\uniiwrHeidelberg}\\
\end{center}
\begin{bibunit}[apsrev4-2]
\section{Method Related Details}

\subsection{Derivation of Lower Resolution Coarse-Grained Models}
\label{si:derivation-lower-resolution-cg-models}
\begin{figure}[H]
    \centering
    \includegraphics[width=\linewidth]{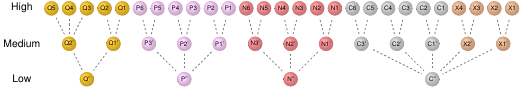}
    \caption{Bead hierarchy for the three coarse-grained (CG) resolutions. The high-resolution model, which is based on the Martini3 force field,\citep{Souza2021} contains 27 distinct bead classes per bead size; accounting for charges on the Q-beads increases this to 32 classes per bead size. The dashed lines indicate how bead types are grouped into medium- and low-resolution types.}
    \label{fig:si:bead-hierarchy}
\end{figure}

For our multi-level Bayesian optimization (BO) approach, we define several levels of coarse-grained (CG) resolution.
We apply the same spatial coarse-graining scheme---mapping atoms to beads---but achieve different resolutions by varying the number of transferable bead types.
Each resolution spans the same chemical space (CS) region but at a different level of chemical detail.
As our high-resolution model, we use the Martini3 force field,\citep{Souza2021} excluding bead labels and water and divalent ion beads.
This high-resolution model includes five Q-beads, six P-beads, six N-beads, six C-beads, and four X-beads for each of the three bead sizes.
Because Q-beads carry either a positive or negative charge, we obtain 32 classes per bead size and 96 bead types in total.
A detailed description of the bead properties can be found in \citeauthor{Souza2021}\citep{Souza2021}
Within our framework, the set of available bead types fully determines a CS resolution.
All small molecules that can be assembled from these bead types are part of the corresponding CS resolution (see Section~\ref{si:molecular-graph-enumeration}).
We hierarchically combine bead types to construct lower-resolution CG models.
Figure~\ref{fig:si:bead-hierarchy} illustrates the hierarchical relationships among bead types for a single bead size; the same scheme applies across all bead sizes and for both positive and negative Q-bead charges.

\begin{figure}[H]
    \centering
    \includegraphics[width=\linewidth]{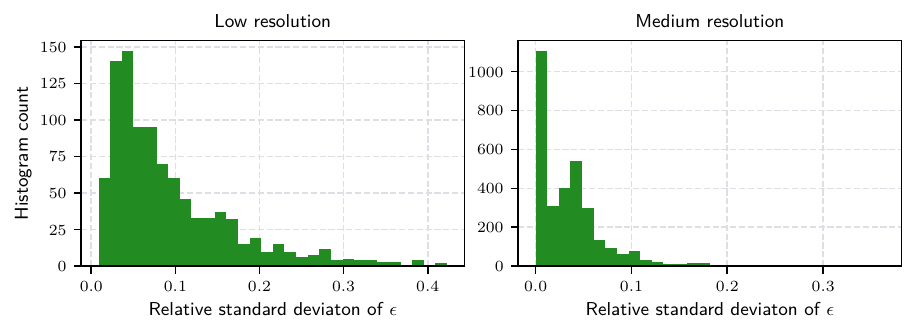}
    \caption{Relative standard deviations of the Lennard-Jones parameter $\epsilon$\citep{GromacsNonbondedInteractions} from the parameter averaging for the medium- and low-resolution CG models. Variations in the averaged low-resolution parameters are higher, as expected by the broader range of bead types.}
    \label{fig:si:mapping-epsilon-std}
\end{figure}

The Martini3 model defines only general non-bonded interactions; intra-molecular interactions must be parameterized individually for each molecule.
While charged interactions are described using a standard electrostatic potential, the parameterized Lennard-Jones interactions are specific to Martini3.
Because our high-resolution model is based on Martini3, we could directly adopt its parameters in our simulations.
Lennard-Jones parameters for the lower-resolution models were obtained by averaging the corresponding high-resolution parameters.
For example, the parameters for the P'' bead were calculated by averaging over the P1 to P6 beads in the Martini3 model.
Since bead sizes remain the same across resolutions, only the interaction strength of the Lennard-Jones potential (denoted as $\epsilon$) is effectively averaged.\citep{GromacsNonbondedInteractions}
Figure~\ref{fig:si:mapping-epsilon-std} shows the relative standard deviation of the averaged $\epsilon$ parameters for both the medium- and low-resolution models.
At the low and medium resolutions, the relative standard deviation is below 10\% for approximately 67\% and 94\% of the parameters, respectively.
These relatively slight variations suggest that the combination of bead types is reasonable.
As expected, the low-resolution model exhibits greater variability in the averaged parameters.

\subsection{Molecular Graph Enumeration}
\label{si:molecular-graph-enumeration}
Since Martini3 only characterizes non-bonded interactions between beads, intra-atomic bonded forces must be determined individually for each molecule.
However, starting from atomistic structures and mapping them to CG representations is not feasible across large regions of chemical space.
Therefore, we generated molecules directly at the CG level, which has fewer combinatorial possibilities due to the reduced resolution.
Although we bypassed the atomistic level initially, backmapping remained possible based on the properties of the beads.\citep{Mohr2022}

To enumerate CG small-molecule CS, we applied several simplifications and assumptions.
We treated small molecules as graphs with fixed, bead-size-dependent bond lengths (listed in Table~\ref{tab:si:bond-lengths}).
Furthermore, we neglected angle and dihedral interactions, a choice justified by the small size of the molecules.
While using fixed bond lengths ignored some chemical details, we assumed that the bead identities and their overall arrangement were more critical than minor variations in bond lengths.
We further restricted the small-molecule CS to molecules containing up to four CG beads, corresponding to atomistic molecules with up to 16 heavy atoms.

\begin{table}[b]
    \centering
    \caption{Bead-size–dependent bond lengths. These bond lengths are used to convert molecular graphs into CG representations suitable for simulation. R, S, and T correspond to regular, small, and tiny beads.}
    \begin{tabular}{ccc}
        Bead size 1 & Bead size 2 & Bond length / nm \\
        \hline
        T & T & 0.29 \\
        T & S & 0.31 \\
        S & S & 0.33 \\
        T & R & 0.33 \\
        S & R & 0.35 \\
        R & R & 0.38 \\
    \end{tabular}
    \label{tab:si:bond-lengths}
\end{table}

The main challenge in enumerating all graphs with a given number of nodes was avoiding duplicates, i.e., isomorphic graphs.
To address this, we separated bead-type enumeration from bond/topology generation.
Enumerating unique sets of bead types effectively reduced to a combination-with-repetition problem, which was straightforward to solve using standard tools, such as those available in Python.\citep{PythonItertools}
Since many bead-type combinations shared the same unique bond configurations, we could reuse generated topologies across multiple bead-type sets.
For example, the bead sets \{{\sffamily C1, C1, C2, C3}\} and \{{\sffamily C2, C2, C1, C4}\} shared the same set of bond configurations.

The generation of topologies for a given set of bead types corresponds to generating all non-isomorphic graphs.
We employed a two-step filtering strategy to avoid computationally infeasible exhaustive graph isomorphism checks.
First, we enumerated all possible topologies using an adjacency matrix representation.
To check whether a graph had been previously generated, we created a simple vectorial fingerprint based on node degrees and the number of bonds per bead type.
This fingerprint was identical for isomorphic graphs but not necessarily unique for non-isomorphic ones.
By storing previously generated graphs in a dictionary-like structure keyed by these fingerprints, we restricted expensive graph isomorphism checks to graphs sharing the same fingerprint.

Overall, the separation of bead and bond generation, the reuse of bond configurations across equivalent bead sets, and the two-stage isomorphism check allowed us to generate unique molecular topologies efficiently.
Although we focused on molecules with up to four beads, the molecule generation algorithm could enumerate much larger molecular graphs.

The exact numbers of generated molecular graphs with up to four beads for the three CS resolutions with 15, 45, and 96 available bead types were 89960, 6742680, and 136870880, respectively.

\subsection{Autoencoder for Molecular Graph Embedding}
\label{si:rae-graph-embedding}

\begin{figure}[t]
    \centering
    \includegraphics[width=0.75\linewidth]{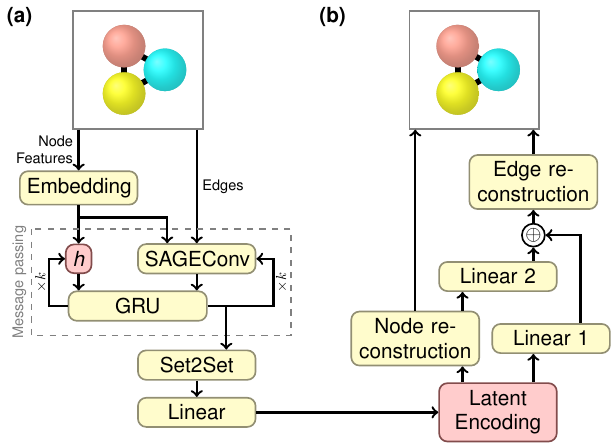}
    \caption{Overview of our regularized autoencoder (RAE) architecture. (a) The molecular graph encoder includes a node embedding layer, a SAGEConv- and GRU-based\citep{Hamilton2017, Cho2014, Gilmer2017} message passing, and Set2Set\citep{Vinyals2015} aggregation. (b) The decoder comprises four feedforward neural networks (FNNs) that sequentially reconstruct the node features and then the adjacency matrix, using the reconstructed nodes and the latent encoding. Although the decoder is not permutation invariant, the loss function is. The symbol $\oplus$ denotes concatenation. Parameter details are listed in Table~\ref{tab:si:rae-hyperparameters}. The architecture is inspired by \citeauthor{Mohr2022}\citep{Mohr2022}}
    \label{fig:si:rae-architecture}
\end{figure}

We encoded all previously enumerated molecules into a five-dimensional latent representation to facilitate molecular optimization.
This representation enabled distance calculations between molecules.
Assuming a sufficiently smooth molecular encoding, this distance was used to measure molecular similarity.
We employed a graph neural network (GNN)-based regularized autoencoder (RAE)\citep{Ghosh2020} model to generate a latent encoding for each CS resolution separately.
The general RAE structure was identical for each resolution.
However, we used different numbers of learnable parameters, with autoencoders for higher CS resolutions containing more parameters.
We implemented the autoencoder using the PyTorch and PyTorchGeometric libraries.\citep{Paszke2019, Fey2019}
The general architecture was inspired by \citeauthor{Mohr2022}\citep{Mohr2022}

The encoder part of the autoencoder consisted of a SAGEConv- and GRU-based\citep{Hamilton2017, Cho2014, Gilmer2017} GNN with $k=4$ message-passing steps.
The GNN-based architecture ensured a permutation-invariant encoding.
Figure~\ref{fig:si:rae-architecture}a illustrates the architecture of the graph encoder, and Table~\ref{tab:si:rae-hyperparameters} lists the hyperparameters for each component.

The decoder part of the autoencoder consisted of four feedforward neural networks (FNNs).
First, the node features were decoded from the latent space representation.
Next, the decoded nodes were passed through another FNN and concatenated with an additional decoded representation of the latent space.
Finally, the concatenated result was passed through a final FNN to generate the triangular part of the symmetric adjacency matrix of the graph.
Figure~\ref{fig:si:rae-architecture}b shows the graph decoder architecture, and Table~\ref{tab:si:rae-hyperparameters} provides the corresponding hyperparameters.
Since the nodes and adjacency matrix were decoded each in a single step, the decoder was not permutation invariant.
However, we observed better autoencoder performance with a one-shot decoding than permutation-invariant decoding strategies.

\newcommand{\spacedhline}{\hline\\[-1.5ex]}
\begin{table}
    \centering
    \caption{Architecture and parameter details for the RAE molecular encoder. ``Layer name'' refers to the labels shown in Figure~\ref{fig:si:rae-architecture}, and ``Model'' describes the corresponding applied function. $x$ denotes the input to each layer, and $\alpha$ corresponds to the LeakyReLU activation function.\citep{Xu2015} All linear layers included a bias term. The third to fifth columns report the number of parameters for the linear transformations, excluding bias terms. When all three resolutions used the same transformation, values are provided only once for clarity.}
    \begin{tabular}{p{3.5cm}p{4cm}p{3cm}p{3cm}p{3cm}}
    & & \multicolumn{3}{c}{\textbf{Parameters for Encoding at Resolution}} \\
    \textbf{Layer Name} & \textbf{Model} & \makecell{\textbf{Low}} & \makecell{\textbf{Medium}} & \makecell{\textbf{High}} \\
    \hline
    \spacedhline
    Embedding & $\alpha(W_{2}\ \alpha(W_{1}\ x))$ & \makecell{$W_1: 10\times 128$ \\ $W_2: 128\times 64$} & \makecell{$W_1: 19\times 256$ \\ $W_2: 256\times 64$} & \makecell{$W_1: 33\times 512$ \\ $W_2: 512\times 64$}\\
    \spacedhline
    SAGEConv & $W_1\ x_i + W_2\ \text{mean}_{j\in\mathcal{N}(i)} x_j$ &  & \makecell{$W_1: 64\times 64$ \\ $W_2: 64\times 64$} & \\
    \spacedhline
    GRU & GRU(x, h) &  & \makecell{dim. 64} &  \\
    \spacedhline
    Linear & $\alpha(W_{2}\ \alpha(W_{1} x))$ & & \makecell{$W_1: 128\times 64$ \\ $W_2: 64\times 5$} & \\
    \spacedhline
    Node reconstruction & $W_{3}\ \alpha(W_{2}\ \alpha(W_{1} x))$ & \makecell{$W_1: 5\times 1024$ \\ $W_2: 1024\times 512$ \\ $W_3: 512\times 44$} & \makecell{$W_1: 5\times 1024$ \\ $W_2: 1024\times 1024$ \\ $W_3: 1024\times 80$} & \makecell{$W_1: 5\times 2048$ \\ $W_2: 2048\times 1024$ \\ $W_3: 1024\times 136$}\\
    \spacedhline
    Linear 1 & $W_{2}\ \alpha(W_{1} x)$ & & \makecell{$W_1: 5\times 512$ \\ $W_2: 512\times 512$} & \\
    \spacedhline
    Linear 2 & $W_{2}\ \alpha(W_{1} x)$ & \makecell{$W_1: 10\times 128$ \\ $W_2: 128\times 32$} & \makecell{$W_1: 19\times 256$ \\ $W_2: 256\times 32$} & \makecell{$W_1: 33\times 512$ \\ $W_2: 512\times 32$} \\
    \spacedhline
    Edge reconstruction & $W_{3}\ \alpha(W_{2}\ \alpha(W_{1} x))$ & & \makecell{$W_1: 640\times 512$ \\ $W_2: 512\times 512$ \\ $W_3: 512\times 6$} & \\
    \spacedhline
    \end{tabular}
    \label{tab:si:rae-hyperparameters}
\end{table}

\begin{table}[htb]
    \centering
    \caption{Reconstruction accuracies for edges, bead classes, sizes, and charges across all three encoded CS resolutions. For the low-resolution model, accuracies were evaluated over all molecules, while for the medium- and high-resolution models, they were computed on samples of $10^6$ molecules.}
    \begin{tabular}{lp{2.5cm}p{2.5cm}p{2.5cm}p{2.5cm}}
    \multirow{2}{*}{\textbf{Resolution level}} & \multicolumn{3}{l}{\textbf{Reconstruction accuracies}} \\
     & \textbf{Edges} & \textbf{Bead classes} & \textbf{Bead sizes} & \textbf{Bead charges} \\
    \hline
    Low & 0.976 & 0.989 & 0.993 & 1.000\\
    Medium & 0.994 & 0.999 & 0.991 & 1.000 \\
    High & 0.987 & 0.979 & 0.986 & 0.998\\
    \end{tabular}
    \label{tab:si:reconstruction-accuracies}
\end{table}

For training, we used an Adam optimizer\citep{Kingma2014} with the standard parameters $\beta_1=0.9$ and $\beta_2=0.999$ and a learning rate of \SI{1e-3}{}.
The loss function
\begin{equation}
    \mathcal{L} = \mathcal{L}\textsubscript{reconstruction} + \lambda\textsubscript{weights} \mathcal{L}_{w-\text{decoder}} + \lambda\textsubscript{latent} \| z \|_2
\end{equation}
consisted of a reconstruction term $\mathcal{L}\textsubscript{reconstruction}$, a decoder weight regularization term $\mathcal{L}_{w-\text{decoder}}$ with prefactor $\lambda\textsubscript{weights}=\SI{1e-5}{}$, and a L2-regularization of the latent space $z$ with prefactor $\lambda\textsubscript{latent}=\SI{1e-4}{}$.
The reconstruction loss $\mathcal{L}\textsubscript{reconstruction}$ consisted of three cross-entropy loss terms for the categorical node features, a mean squared error term for the octanol-water partitioning free energy, and a binary cross-entropy for the triangular adjacency matrix.
We minimized $\mathcal{L}\textsubscript{reconstruction}$ over all possible graph permutations to achieve a permutation-invariant reconstruction loss.
Because we considered only graphs with up to four beads, evaluating all permutations was computationally feasible.
For larger molecules, however, this approach becomes prohibitively expensive.
We used a training batch size of 16,384 molecular graphs.
The node features included three categorical properties---class, size, and charge---and the octanol-water partitioning free energy as a continuous property.
All categorical features were one-hot encoded and concatenated with the octanol-water partitioning free energy before encoding.
We used a dummy bead class during node reconstruction to reconstruct molecules with fewer than four beads.
If this dummy class was predicted, the node and its corresponding edges were ignored.

Because our goal was to obtain a numerical encoding of the chemical space without generating new molecules, we discarded the decoder after training.
Nonetheless, we aimed for high reconstruction accuracy, indicating high information content in the encoded representation.
Table~\ref{tab:si:reconstruction-accuracies} reports the reconstruction accuracies for all three CS encodings.

\subsection{Mapping Between Latent Spaces and Neighborhood Calculations}
\label{si:latent-space-mapping-and-neighborhoods}

In our multi-level BO algorithm, we used lower-resolution information as a prior to guide exploration at higher resolutions of CS, which required a mapping between resolutions.
Because we learned the CS representations separately for each CG resolution, the latent vectors could not be directly transferred across resolutions.
However, since we defined a known bead-type mapping between resolutions (see Section~\ref{si:derivation-lower-resolution-cg-models}), we could establish a correspondence between molecules at each resolution and therefore between their latent representations.
For example, the two-bead molecule {\sffamily P4--C5} at high resolution could be mapped to {\sffamily P2'--C3'} at medium resolution and to {\sffamily P''--C''} at low resolution.

In practice, we stored all enumerated molecules in an indexable database, along with their latent representation and a reference to the corresponding next-lower-resolution molecule.
Creating a database index for this reference at both resolution levels made it possible to map molecules quickly between lower and higher resolutions.
As a result, information from lower-resolution Gaussian process (GP) models could be easily transferred to higher-resolution levels.\\

For the optimization at higher resolutions, our multi-level BO algorithm focused on latent space regions with the highest likelihood of favorable $\Delta\Delta G$ values.
In each optimization iteration, the expected improvement (EI) based on the GP model was maximized only within the neighborhoods of the most promising molecules.
This neighborhood-focused maximization at resolution level $l$ (with $l \geq 2$) followed several steps.

First, we selected the top $m$ molecules $X_{l-1,\text{top-}m}$ with the lowest $\Delta\Delta G$ values from the next-lower resolution level $l-1$: $X_{l-1,\text{top-}m} = \arg\max_m \Delta\Delta G(x)$ for $x \in X_{l-1}$.
The hyperparameter $m$, which controlled the balance between optimization accuracy and computational cost, was set to 30.

Next, we identified the set of molecules $\tilde{P}_l$ at level $l$ that corresponded to these $m$ top-ranked molecules at level $l-1$: $\tilde{P}_l = \{x \in \mathcal{X}_l | \mathcal{M}(x) \in X_{l-1,\text{top-}m}\}$.
Using a cell list built over the latent space, we then assembled the set $P_l$ of all molecules at level $l$ that were located in the same or an adjacent cell as any molecule in $\tilde{P}_l$: $P_l = \{x \in \mathcal{X}_l | \exists z \in \tilde{P}_l, d(x,z) \le 1\}$, where $d(x, z)$ was the cell-list-based distance metric.

Because the latent space distribution was approximately Gaussian---due to the $L_2$ loss applied during autoencoder training---we determined the cell sizes in the cell list based on a Gaussian distribution function. Finally, the EI was maximized over all points $x \in P_l$.

This neighborhood-restricted EI maximization allowed the algorithm to efficiently concentrate on promising regions of chemical space while maintaining computational tractability.

\subsection{Scoring of Water- or Interface-Localizing Molecules}
\label{si:score-calculation}

To estimate a molecule's lipid bilayer demixing behavior, we calculated the free-energy difference between inserting the molecule into a ternary versus a pure DLiPC bilayer, following the approach of \citeauthor{Centi2020}\citep{Centi2020}
We focused solely on the free-energy difference between the bilayer centers to avoid the computational expense of computing potential of mean force (PMF) profiles in both bilayers.
Prior observations that molecules influencing bilayer mixing tend to localize near the bilayer center justified this simplification.\citep{Centi2020}
Specifically, we computed the free-energy difference $\Delta\Delta G = \Delta G\textsubscript{center} - \Delta G\textsubscript{DLiPC}$ between positioning the molecule at the center of the ternary and the center of the DLiPC bilayer.
However, this measure was meaningful only for molecules localizing at the bilayer center.
To account for localization behavior, we combined $\Delta\Delta G$ with a score $S$ that penalized molecules preferring bulk water or the bilayer interface over the center.
The penalty score was defined by the following conditional equation:
\begin{equation}
\label{equ:si:score-calculation}
    S = \begin{cases}
    0.5 + \min(\Delta G\textsubscript{water} - \Delta G\textsubscript{center}, 25) \dfrac{2}{25}, & \text{if}\ \Delta G\textsubscript{water} > \Delta G\textsubscript{center} \\
    \min(\Delta G\textsubscript{interface} - \Delta G\textsubscript{center}, 3) \dfrac{0.5}{3}, & \text{else if}\ \Delta G\textsubscript{interface} > \Delta G\textsubscript{center} \\
    0, & \text{else.}
    \end{cases}
\end{equation}
This equation scaled the free-energy differences between the center and the interface or water phase to values between \SIrange{0}{2.5}{\kilo\calorie\per\mol}.
The scaling and thresholds were chosen empirically, informed by typical $\Delta G\textsubscript{center}$, $\Delta G\textsubscript{interface}$, and $\Delta G\textsubscript{water}$ values and by expected $\Delta\Delta G$ patterns reported in prior simulations.\citep{Centi2020}
The formulation assumed that most $\Delta G\textsubscript{water} - \Delta G\textsubscript{center}$ values fell between \SIrange{0}{25}{\kilo\calorie\per\mol}, and $\Delta G\textsubscript{interface} - \Delta G\textsubscript{center}$ values between \SIrange{0}{3}{\kilo\calorie\per\mol}.

By incorporating the score $S$, we made the free-energy landscape more meaningful in regions of chemical space where molecules did not localize at the bilayer center.
Without this adjustment, raw $\Delta\Delta G$ values would have produced an uninformative landscape over large regions, hindering optimization within CS.

\subsection{Simulation Parameters}
\label{si:simulation-parameters}
Each molecule evaluation within our active learning loop involved up to four thermodynamic integration (TI) calculations.
For each of these calculations, we employed 26 or 36 linearly spaced $\lambda$-steps for uncharged and charged molecules, respectively.
Simulations were conducted using GROMACS\citep{Abraham2015, Pall2020} in an automated high-throughput workflow, applying a consistent configuration across all runs.
Initially, each system underwent a setup step, an energy minimization, and an equilibration.
The same equilibrated structure was subsequently used for all $\lambda$-step simulations.
For the system setup, the small CG molecule was placed in the respective environment---either bulk water or one of the lipid bilayer systems.
Energy minimization was performed using the steepest descent method with 30,000 optimization steps.
For the equilibration, we used an integration time step of \SI{10}{\femto\second} (in reduced CG units) over 40,000 integration steps.
The subsequent $\lambda$-step simulations used a \SI{20}{\femto\second} integration time step with the following number of integration steps, depending on the system:
\begin{center}
    \centering
    \begin{tabular}{rr}
        Bulk water: & 400,000 \\
        DLiPC bilayer center: & 400,000 \\
        Ternary bilayer center: & 900,000 \\
        Ternary bilayer interface: & 1,200,000 \\
    \end{tabular}
\end{center}
Different step counts were used to balance computational efficiency with adequate convergence, recognizing, for example, the greater heterogeneity of the ternary bilayer compared to the pure DLiPC bilayer.
Due to the conditional construction of the score, simulations at the ternary bilayer interface and the DLiPC center were skipped for molecules that did not insert into the bilayer. As shown in Table \ref{tab:si:computational_load}, this results in a lower average computational load per simulation at low resolution than at both higher resolutions. For the $\lambda$ simulations, we achieved an average simulation performance of \SI{4.3}{\micro\second}/day or \SI{2.16e8}{} steps/day.

\begin{table}[ht]
    \centering
    \caption{Total number of simulations and simulation steps per resolution level including all equilibration and $\lambda$ simulations. Due to the conditional construction of the score and funnel-like optimization, lower resolution evaluations have a larger average computational load.}
    \begin{tabular}{lcc}
        Resolution & \# simulations & \# total simulation steps \\
        \hline
        Low        & 91 (29.2\%)     & \num{2.0e10} (23.3\%) \\
        Medium     & 148 (47.4\%)    & \num{4.5e10} (51.4\%) \\
        High       & 73 (23.4\%)     & \num{2.2e10} (25.3\%) \\
    \end{tabular}
    \label{tab:si:computational_load}
\end{table}

Both the equilibration and $\lambda$-step simulations employed a leap-frog stochastic dynamics integrator at a temperature of \SI{305}{\kelvin}, with an inverse friction constant of \SI{2}{\pico\second}.
Pressure was maintained at \SI{1}{\bar} using a semi-isotropic C-rescale barostat\citep{Bernetti2020} with a relaxation time constant of \SI{4}{\pico\second} and a compressibility of \SI{3e-4}{\per\bar}.
Coulomb interactions beyond a \SI{1.2}{\nano\meter} cutoff were treated using the reaction field method, while Lennard-Jones interactions were truncated and shifted at \SI{1.2}{\nano\meter}.
To minimize artifacts, we set the GROMACS parameters \texttt{verlet-buffer-tolerance} to $-1$ and \texttt{rlist} to $1.4$.\citep{Kim2023}
To restrain the molecule's $z$-position within the lipid bilayer, we applied an umbrella potential with a force constant of \SI{500}{\kilo\joule\per\mol\per\nano\meter\squared}.
The restraining reference was defined over a cylindrical bilayer region with a radius of \SI{1.5}{\nano\meter} around the inserted molecule (\texttt{pull-coord1-geometry = cylinder}).

To measure or restrain the number of DPPC-DIPC contacts in the ternary bilayer, we used the colvars module in GROMACS.\citep{Fiorin2013}
For each leaflet, a collective variable was defined using the coordination number (\texttt{coordNum} option) between the first C1 beads of the two phospholipids, with a cutoff distance of \SI{1.1}{\nano\meter}.
To improve performance, we set the colvars pair list \texttt{tolerance} to 0.001 and the pair list frequency to 100.
The collective variable was restrained using a harmonic potential with a force constant scaling of 1.

\section{Result Related Details}

\subsection{Analysis of the Learned Molecular Latent Space Representations}
\label{si:latent-space-analysis}
All molecules in the chemical space are mapped to a five-dimensional latent representation to facilitate molecule optimization.
Navigation within this continuous latent space is more tractable than in the discrete molecular space.
The latent representations are learned independently for each resolution level using a GNN-based RAE (see Section~\ref{si:rae-graph-embedding}).
Visualizing these five-dimensional latent spaces in an informative manner is inherently challenging.
A two-dimensional principal component analysis (PCA) projection provides an overview of the latent structure (see Figure \ref{fig:latent-space-points} in the main text).
However, as all five latent dimensions carry meaningful information, the PCA projection captures only a limited portion of the variance, inevitably discarding essential details.
To provide deeper insights into the learned representations, four out of the ten combinatorially possible two-dimensional projections are visualized for each resolution level, colored by various molecular properties.
Figure~\ref{fig:si:latent-space-props-octanol} shows the distribution of the summed octanol–water transfer free energies of individual beads, $\Delta\Delta G_{\text{octanol}\rightarrow\text{water}}$, across the latent space.
A clear correspondence between the latent space structure and the summed $\Delta\Delta G_{\text{octanol}\rightarrow\text{water}}$ is observed. This behavior is expected given that these values serve as inputs and reconstruction targets for the autoencoder.
Additional visualizations for total molecular charge, molecular weight, number of nodes, and number of edges are presented in Figures~\ref{fig:si:latent-space-props-charge}, \ref{fig:si:latent-space-props-weight}, \ref{fig:si:latent-space-props-nodes}, and \ref{fig:si:latent-space-props-edges}, respectively. For medium- and high-resolution levels, these visualizations are based on random latent space samples of 100,000 molecules.
All figures reveal discernible structural patterns associated with the underlying molecular properties.

\newcommand{\latentSpacePropsWidth}{0.93\linewidth}
\begin{figure}[H]
    \centering
    \includegraphics[width=\latentSpacePropsWidth]{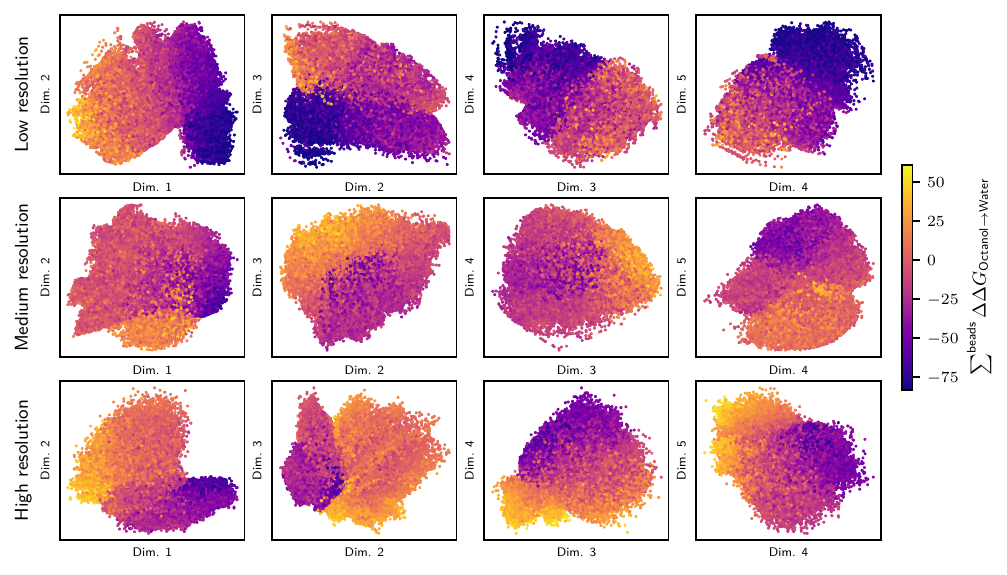}
    \caption{Learned latent spaces for the three resolution levels, with points colored by sum over octanol-water transfer free energies for individual beads. For the medium and high resolution, 100,000 molecules were randomly sampled from the full latent space.}
    \label{fig:si:latent-space-props-octanol}
\end{figure}
\begin{figure}[H]
    \centering
    \includegraphics[width=\latentSpacePropsWidth]{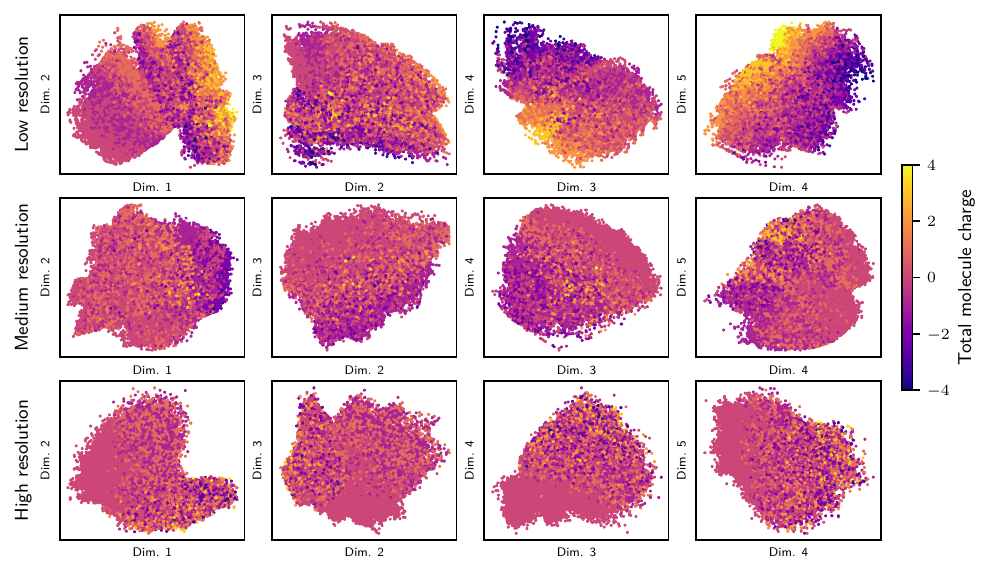}
    \caption{Learned latent spaces for the three resolution levels, with points colored by the total molecule charge. For the medium and high resolution, 100,000 molecules were randomly sampled from the full latent space.}
    \label{fig:si:latent-space-props-charge}
\end{figure}
\begin{figure}[H]
    \centering
    \includegraphics[width=\latentSpacePropsWidth]{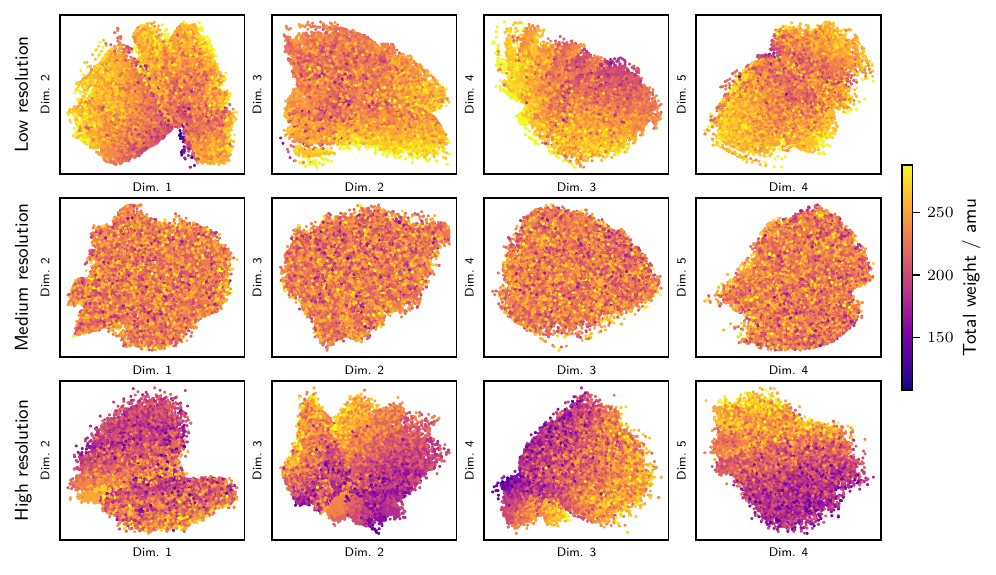}
    \caption{Learned latent spaces for the three resolution levels, with points colored by the total molecule weight. For the medium and high resolution, 100,000 molecules were randomly sampled from the full latent space.}
    \label{fig:si:latent-space-props-weight}
\end{figure}
\begin{figure}[H]
    \centering
    \includegraphics[width=\latentSpacePropsWidth]{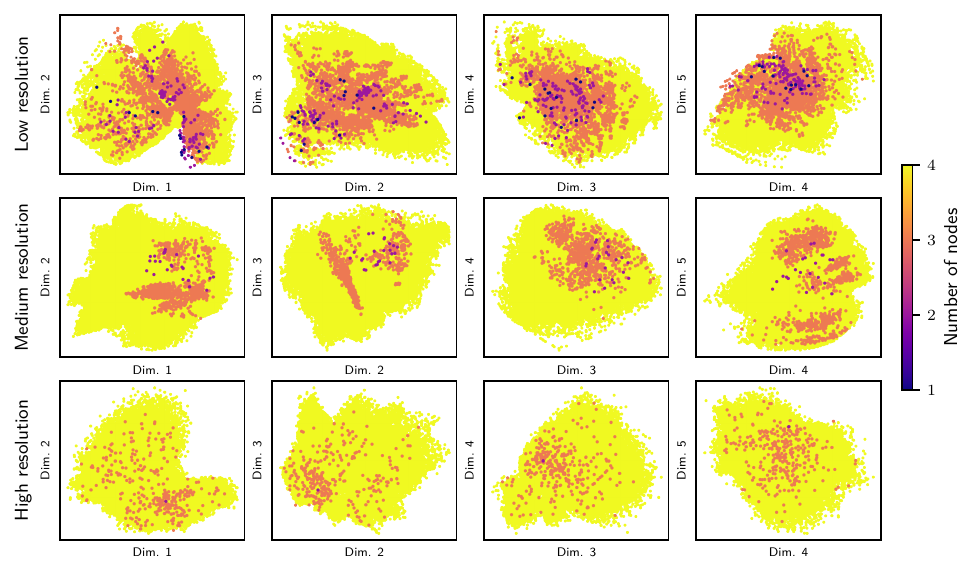}
    \caption{Learned latent spaces for the three resolution levels, with points colored by the number of nodes per molecule. For the medium and high resolution, 100,000 molecules were randomly sampled. Scatter points are drawn in decreasing node count order to enhance the visibility of molecules with low node counts.}
    \label{fig:si:latent-space-props-nodes}
\end{figure}
\begin{figure}[H]
    \centering
    \includegraphics[width=\latentSpacePropsWidth]{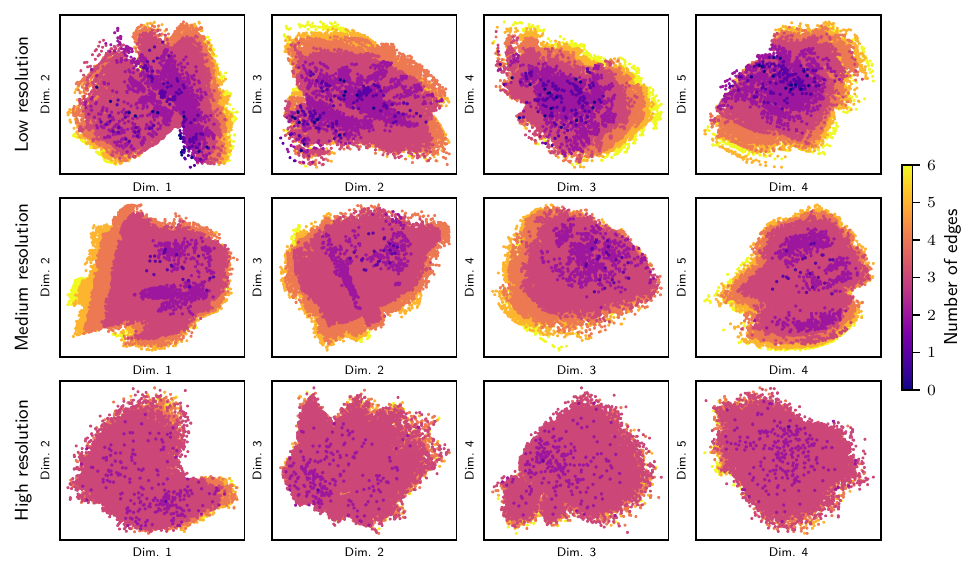}
    \caption{Learned latent spaces for the three resolution levels, with points colored by the number of edges per molecule. For the medium and high resolution, 100,000 molecules were randomly sampled.  Scatter points are drawn in decreasing node count order to enhance the visibility of molecules with low node counts.}
    \label{fig:si:latent-space-props-edges}
\end{figure}

\subsection{Lowest-Resolution Model Prior}
\label{si:additivity-assumption}
\begin{figure}
\includegraphics[width=\linewidth]{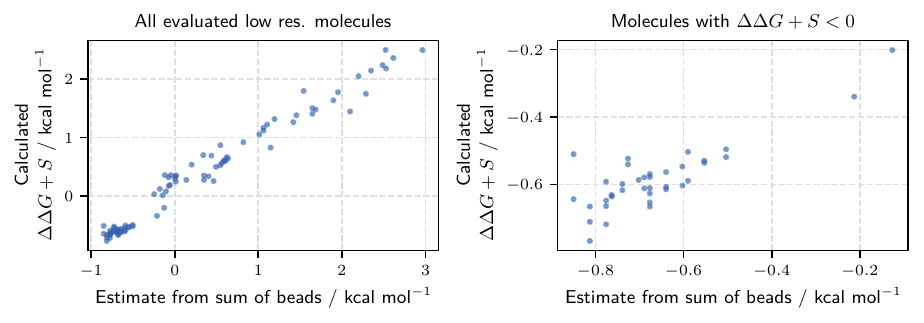}
\caption{Comparison of calculated molecule $\Delta\Delta G+S$ values with estimates based on a linear combination of the individual bead $\Delta\Delta G+S$ contributions. The left panel presents results for all molecules evaluated at the low-resolution level. Many molecules do not localize near the bilayer center, resulting in positive $S$ values. The right panel includes only molecule results with negative $\Delta\Delta G$ (i.e., $S=0$). Both panels show a clear correlation between the simulation results and the bead-based estimation.}
\label{fig:si:prior-vs-result}
\end{figure}

In our multi-level BO algorithm, the GP model at resolution level $l-1$ serves as the mean prior for the GP model at resolution level $l$.
However, no lower-level model is available at the lowest-resolution level ($l=1$).
Rather than using a constant prior, we employ a free-energy estimate based on the additive contribution of individual CG bead free energies:
\begin{equation}
    \widehat{(\Delta\Delta G + S)}\textsubscript{molecule} = \beta \sum^{\text{beads}} (\Delta\Delta G + S)\textsubscript{bead}.
\end{equation}
While correlations between beads influence a molecule's free-energy result, the additivity assumption provides a reasonable approximation.
Figure~\ref{fig:si:prior-vs-result} compares $\Delta\Delta G+S$ values obtained from full molecule simulations to those estimated by the linear combination of bead contributions.
The results show a strong correlation, supporting the validity of the additive assumption as a prior for the GP model.
The scaling parameter $\beta$ was determined by fitting the simulated free energies of the 50 initialization molecules to their corresponding bead sums, yielding $\beta = 0.63$.

\subsection{Evaluation of the \texorpdfstring{$\Delta\Delta G$}{ΔΔG} Standard Deviation}
\label{si:free-energy-standard-deviation}

\begin{figure}
\includegraphics[width=\linewidth]{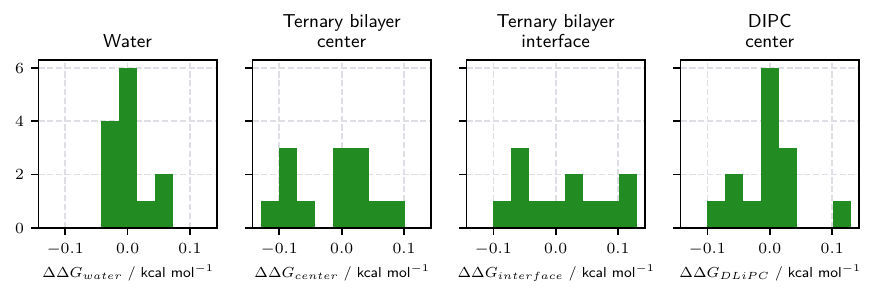}
\caption{Histograms of differences between two repeated calculations of $\Delta G$ values for 14 molecules in water, at the ternary bilayer center and interface, and the DLiPC bilayer center.}
\label{fig:si:free-energy-standard-deviation}
\end{figure}
The GP model with a radial basis function (RBF) kernel (described in Section~\ref{sec:methods:multi-level-bo} in the main text) has two parameters: the kernel lengthscale, $\xi$, and the data noise, $\sigma\textsubscript{n}$.
Both parameters can be obtained via maximum likelihood estimation during the GP regression process.
However, to improve the robustness of the lengthscale estimation, we fixed the value of $\sigma_\text{n}$ based on the observed variability in the computed free energies.
This was achieved by performing duplicate free-energy calculations for 14 molecules across the four different environments (water, ternary bilayer interface and center, and DLiPC bilayer center), and using the resulting differences to estimate the standard deviations of $\Delta G$, assuming a Gaussian distribution of errors.
This approach was also used to determine the number of molecular dynamics (MD) integration steps for each system (see \ref{si:simulation-parameters}) required to achieve an acceptable accuracy.
Figure~\ref{fig:si:free-energy-standard-deviation} presents histograms of the observed $\Delta G$ differences between paired simulations.
Table~\ref{tab:si:free-energy-standard-deviations} summarizes the corresponding standard deviations.
Based on these results and equation~\ref{equ:si:score-calculation}, we set $\sigma\textsubscript{n} = \SI{0.05}{\kilo\calorie\per\mol}$.

\begin{table}[H]
    \centering
    \caption{Standard deviations of $\Delta G$ for different systems.}
    \label{tab:si:free-energy-standard-deviations}
    \begin{tabular}{lc}
        \toprule
        System & $\sigma_{\Delta G}$ / kcal mol$^{-1}$ \\
        \midrule
        Water & 0.056 \\
        Ternary bilayer center & 0.051 \\
        Ternary bilayer interface & 0.048 \\
        DLiPC bilayer center & 0.034 \\
        \bottomrule
    \end{tabular}
\end{table}

\subsection{Best obtained molecules from the low and medium resolution optimization}
\label{si:best-low-medium-molecules}

\begin{figure}[H]
    \centering
    \includegraphics[width=\linewidth]{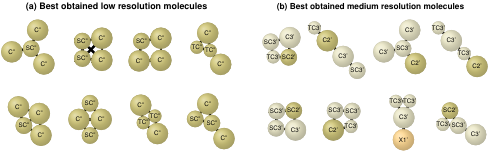}
    \caption{Highest-performing molecules from the (a) low- and (b) medium-resolution models. While the molecules exhibit diverse topologies, they consist solely of {\sffamily C"} beads at low resolution, and primarily {\sffamily C2'} and {\sffamily C3'} beads at medium resolution—with one exception involving an {\sffamily X1'} bead. Notably, no {\sffamily C1'} beads are present, despite their mapping to the same {\sffamily C"} type at low resolution. Although the low-resolution results already reveal relevant chemical features, incorporating higher-resolution models provides additional chemical detail. According to \citeauthor{Barnoud2014}\citep{Barnoud2014}, the presence of {\sffamily C1'} beads—or their corresponding higher-resolution types {\sffamily C1}, {\sffamily C2}, and {\sffamily C3}—is expected to enhance lipid mixing. This insight is only accessible through the integration of higher-resolution models into the multi-level optimization.}
    \label{fig:si:top-molecules-res01}
\end{figure}

\subsection{System Composition for Direct Demixing Analysis}
\label{si:demixing-simulation-composition}
During the multi-level BO, we do not directly calculate the phospholipid demixing behavior of molecules.
Instead, we estimate this behavior based on the free-energy difference of inserting a molecule into a pure DLiPC bilayer versus a ternary bilayer.
This approach significantly reduces the computational cost, as converging accurate free-energy differences requires substantially less simulation time than directly observing demixing, which is hampered by significant fluctuations in mixing behavior.
However, to validate the predictions from the optimization workflow, it is necessary to assess the phospholipid demixing behavior through direct simulations.
For these validation simulations, we increase the bilayer area from $\SI{6}{\nano\meter} \times \SI{6}{\nano\meter}$, as used in the free-energy calculations (see Section~\ref{sec:methods:estimating-demixing}), to $\SI{12}{\nano\meter} \times \SI{12}{\nano\meter}$ to improve statistical reliability.
The same DPPC:DLiPC:cholesterol ratio of 7.0:4.7:5.0 is maintained, corresponding to 94 DPPC, 63 DLiPC, and 67 cholesterol molecules.
The target molecule is added to the system at a solute-to-lipid mass ratio of 4.8\%, calculated using CG masses, to evaluate demixing effects.
The total mass of the ternary bilayer is \SI{333180}{\dalton}, derived from $11 \times \SI{72}{\dalton} + 1 \times \SI{54}{\dalton} = \SI{846}{\dalton}$ for DPPC and DLiPC, and $3 \times \SI{72}{\dalton} + 4 \times \SI{54}{\dalton} + 2 \times \SI{36}{\dalton} = \SI{504}{\dalton}$ for cholesterol.
The best-performing molecule from our multi-level BO has a mass of \SI{234}{\dalton}, resulting in 68 molecules being incorporated into the bilayer. For comparison, benzene (\SI{108}{\dalton}) was added at a total of 148 molecules in its corresponding simulation.

\subsection{Chemical Neighborhood Sizes Based on Gaussian Process Kernel Lengthscales}
\label{si:chemical-neighborhood-sizes}
We introduced the concept of chemical neighborhoods to characterize the smoothness of the learned chemical space representation.
A chemical neighborhood is defined as a group of molecules exhibiting similar properties with respect to the optimization target.
Larger neighborhoods indicate a smoother free energy or target function landscape across chemical space.
We quantified molecular similarity, and consequently the size of a chemical neighborhood, using the lengthscale parameter of an RBF kernel fitted within a GP regression.
Smooth functions with lower variability are best modeled with larger lengthscales, while highly variable functions require shorter lengthscales.
Thus, larger length scales imply smoother target functions and correspondingly larger chemical neighborhoods.
To estimate the lengthscale $\xi_l$ at a given resolution level $l$, we optimized the negative marginal log likelihood using GPyTorch.\citep{Gardner2018}
The size of a chemical neighborhood was then determined by calculating the average number of neighboring molecules within a latent space distance $d < \alpha \xi_l$, with $\alpha = 0.5$. Smaller values of $\alpha$ imply a stricter similarity criterion, whereas larger values allow for looser similarity within neighborhoods.
The average number of neighbors within this distance was computed across ten independent samples of 30,000 randomly selected molecules.
The resulting neighborhood sizes, visualized in Figure \ref{fig:neighborhood-sizes} of the main text, are summarized in Table~\ref{tab:si:neighborhood-sizes}.
The table also reports the sizes of mapped neighborhoods across different resolution levels.
On average, a molecule at low resolution corresponds to approximately 75 molecules at medium resolution, and a molecule at medium resolution corresponds to about 20 molecules at high resolution.
These numbers arise from the hierarchical nature of our CG models.
The sizes of mapped neighborhoods can be calculated based on these numbers.
For example, for the low-resolution neighborhood, we obtain a size of $248.6 \cdot 75 \approx 18,600$ molecules when mapped to the medium resolution.

\begin{table}[H]
    \centering
    \caption{Chemical neighborhood size results for the three resolution levels. The table includes the average sizes of neighborhoods mapped to higher resolutions (\textit{italic} numbers).}
    \label{tab:si:neighborhood-sizes}
    \begin{tabular}{l>{\raggedleft\arraybackslash}m{3cm}>{\raggedleft\arraybackslash}m{3cm}>{\raggedleft\arraybackslash}m{3cm}}
        \toprule
          & \multicolumn{3}{c}{\textbf{Neighborhood size represented at resolution level}} \\
        \textbf{Resolution level} & Low & Medium & High\\
        \midrule
        Low & 249 &  &  \\
        Medium & \textit{18,700} & 23 &  \\
        High & \textit{378,000} & \textit{468} & 37 \\
        \bottomrule
    \end{tabular}
\end{table}

\subsection{Multi-level Bayesian Optimization with a Toy Model}
\label{si:toy-model}
The choice of initialization points can significantly influence the performance of a BO run.
Therefore, it is common practice to average results over multiple runs to ensure a statistically meaningful comparison between different BO methods.
However, performing multiple optimization runs for our bilayer demixing application is impractical due to the high computational cost of evaluating each molecule.
Here, we employed a simple toy model with an easily evaluable molecule score to facilitate a comparative analysis between our multi-level BO algorithm and standard BO at the high-resolution level across multiple runs.

\begin{figure}[H]
    \centering
    \includegraphics[width=\linewidth]{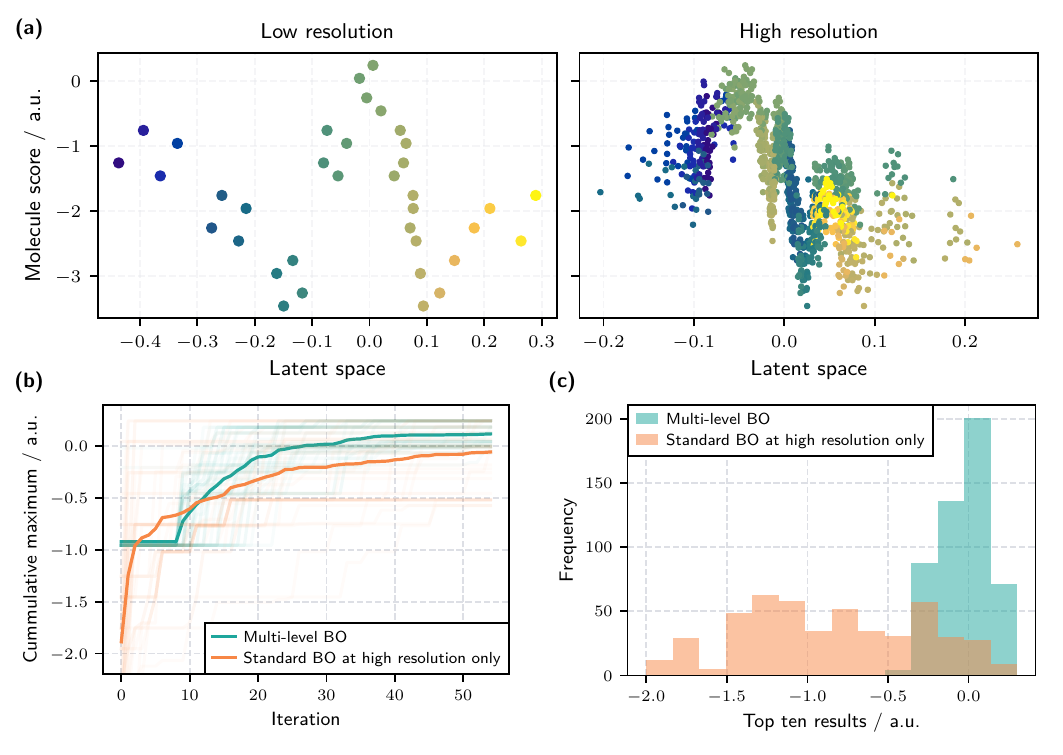}
    \caption{Comparison of standard and multi-level BO on a toy model.
    (a) Learned latent space representation of the discrete chemical space at low (left) and high resolution (right). Molecule scores (equation~\ref{equ:si:toy-model-score}) are shown on the vertical axis. Colors indicate how low-resolution points correspond to their high-resolution counterparts. The high-resolution landscape is notably more complex than the low-resolution one.
    (b) Cumulative best high-resolution molecule scores from 50 runs for standard BO (orange) and multi-level BO (teal), with individual runs (shaded) and their averages (solid). The multi-level approach initially plateaus due to the optimization at low resolution (not shown), but quickly surpasses standard BO after switching to the high-resolution model.
    (c) Histogram of the top ten results per run. Multi-level BO yields a distribution with higher scores and a sharper peak, indicating more consistent convergence toward high-performing solutions compared to standard BO.}
    \label{fig:si:toy-example}
\end{figure}

For this toy model, we considered CG-molecule-like objects with exactly two beads ($n\in{1,2}$), each characterized by two properties, $b_{np}$, that influence the molecule's score.
These properties could, for example, represent charge or polarity.
As a simplification, we assume a non-permutation-invariant molecule score $s$ given by:
\begin{equation}
\label{equ:si:toy-model-score}
    s = -0.5 (b_{11} - b_{21})^2 + 0.5 (b_{11} - 0.3)^2 - b_{12} - (b_{22} - 1)^2
\end{equation}
which we aimed to maximize.
We defined a discrete set of bead values: $b_{n1}\in\{0.00, 0.33, 0.67, 1.00\}$ and $b_{n2}\in\{0.0, 0.25, 0.5, 0.75, 1.0, 1.25, 1.5, 1.75, 2.0\}$, leading to a total of 1,296 possible molecules in this toy example's chemical space.
While direct optimization in this discrete space is relatively simple, we use a learned one-dimensional representation of the chemical space, making the optimization problem more complex and closer to real-world chemical optimizations.
For our multi-level optimization, we used a lower-resolution model with fewer discretization steps ($b_{n1}'\in\{0.00, 1.00\}$ and $b_{n2}'\in\{0.0, 1.0, 2.0\}$), reducing the number of possible molecules to 36.
Figure~\ref{fig:si:toy-example}a visualizes the learned one-dimensional representations and corresponding molecule scores.
Scatter point colors reflect the correspondence between each low-resolution point and its related high-resolution molecules.
Although the scoring function appears relatively simple at low resolution, its structure becomes substantially more complex in the high-resolution model.

We conducted 50 optimization runs using both standard BO and our multi-level BO, following the methodology described in the main text.
Each run was initialized with three randomly selected points, followed by 52 optimization iterations.
Where applicable, we utilized the same hyperparameters, like constraints on the BO kernel lengthscale and noise, for both methods.
Figure~\ref{fig:si:toy-example}b shows the average cumulative best result at the high resolution for both optimization approaches.
Initially, the multi-level method remains constant due to the optimization occurring at the low-resolution level.
As a result, standard BO achieves better performance for a low number of evaluation steps.
However, once the multi-level approach transitions to the high-resolution model, it quickly outperforms standard BO on average.
Figure~\ref{fig:si:toy-example}b shows a histogram of the top ten results from each of the 50 optimization runs.
The distribution of top molecules found by our multi-level BO is shifted towards higher values and exhibits a sharper peak compared to standard BO.
This demonstrates that our method identifies higher cumulative optima and consistently finds multiple solutions near the global optimum.
A similar trend was observed in our bilayer demixing application.

\subsection{Mapping of Evaluated Molecules Between Latent Spaces}
\label{si:latent-space-points-mapping}

\begin{figure}[H]
\includegraphics[width=\linewidth]{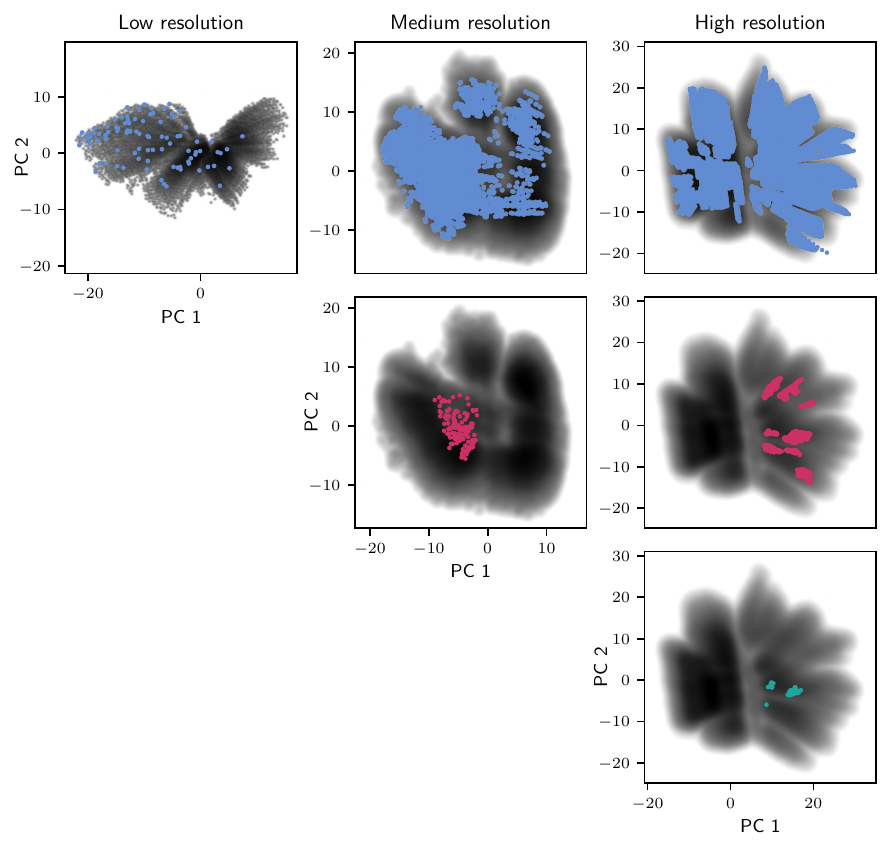}
\caption{Latent space mapping of evaluated molecules across different CS resolutions.
    The diagonal plots display the evaluated molecules at low, medium, and high resolution, respectively (colored points), overlaid on a kernel density estimation (black background) of the full latent representation of CS.
    Off-diagonal plots show the corresponding evaluated points mapped to higher resolutions.
    For example, the first row presents all molecules evaluated at low resolution alongside those at higher resolutions that map down to them.
    The latent spaces at different resolutions are learned independently and cannot be directly mapped onto each other.
    However, the hierarchical design of the CG model enables consistent mapping of individual molecules and thus their corresponding latent space points between resolutions.
    All plots represent 2D PCA projections of the five-dimensional latent space.
    The diagrams illustrate that broad coverage at lower resolutions propagates to broad coverage at higher resolutions.
    The rightmost column of the figure illustrates the funnel-like behavior of the optimization process, where broad coverage at lower resolutions gradually focuses towards promising regions in higher-resolution spaces.}
\label{fig:si:latent-space-mapping}
\end{figure}
\putbib[biblio]
\end{bibunit}

\end{document}